\documentclass[]{interact}

\usepackage{epstopdf}
\usepackage[caption=false]{subfig}

\usepackage[numbers,sort&compress]{natbib}
\bibpunct[, ]{[}{]}{,}{n}{,}{,}

\theoremstyle{plain}

\theoremstyle{definition}

\theoremstyle{remark}

\RequirePackage{array}
\RequirePackage{multirow}
\RequirePackage{makecell}

\usepackage[table]{xcolor}
\definecolor{qianbai}{HTML}{f0f0f4}
\definecolor{yinbai}{HTML}{e9e7ef}

\definecolor{shuangse}{HTML}{e9f1f6}
\definecolor{xuebai}{HTML}{f0fcff}
\definecolor{yingbai}{HTML}{e3f9fd}
\definecolor{yuebai}{HTML}{d6ecf0}

\definecolor{xiangya}{HTML}{fffbf0}
\definecolor{gao}{HTML}{f2ecde}
\definecolor{baifen}{HTML}{fff2df}
\definecolor{yudu}{HTML}{fcefe8}

\definecolor{tubai}{HTML}{f3f9f1}
\definecolor{yaluanqing}{HTML}{e0eee8}
\definecolor{su}{HTML}{e0f0e9}
\definecolor{xiekeqing}{HTML}{bbcdc5}

\definecolor{bailv}{HTML}{d6e9ca}
\definecolor{xiachong}{HTML}{cee4ae}
\definecolor{mengcong}{HTML}{badcad}
\definecolor{yeliu}{HTML}{c1d8ac}
\definecolor{liuse}{HTML}{a8c97f}
\definecolor{liuran}{HTML}{93b881}
\definecolor{shankui}{HTML}{a8bf93}

\definecolor{baiteng}{HTML}{dbd0e6}
\definecolor{danteng}{HTML}{bbc8e6}
\definecolor{tengse}{HTML}{bbbcde}
\definecolor{tengzi}{HTML}{a59aca}

\begin{document}

\articletype{Review Paper}

\title{Computational Approaches of Modelling Human Papillomavirus Transmission and Prevention Strategies: A Systematic Review}

\author{
\name{Weiyi Wang\textsuperscript{a}, Shailendra Sawleshwarkar\textsuperscript{b, c} and Mahendra Piraveenan\textsuperscript{a, d} \thanks{Corresponding author: Mahendra Piraveenan. Email: mahendrarajah.piraveenan@sydney.edu.au}}
\affil{\textsuperscript{a}Modelling and Simulation Research Group, School of Computer Science, Faculty of Engineering, The University of Sydney, Sydney, NSW 2006, Australia; \textsuperscript{b}Sydney Medical School, Faculty of Medicine, The University of Sydney, Sydney, NSW 2006, Australia; \textsuperscript{c}Sydney Infectious Diseases Institute, The University of Sydney; \textsuperscript{d}Charles Perkins Centre, The University of Sydney, John Hopkins Drive, Sydney, NSW 2006, Australia}
}

\maketitle

\begin{abstract}
Human papillomavirus (HPV) infection is the  most common sexually transmitted infection in the world. Persistent  oncogenic  Human papillomavirus infection has been a leading  threat to global health and can lead to serious complications such as cervical cancer. Prevention interventions including vaccination and screening have been proved effective in reducing the risk of HPV-related diseases. In recent decades, computational epidemiology has been serving as a very  useful tool to study HPV transmission dynamics and evaluation of prevention strategies. In this paper, we conduct a comprehensive literature review on state-of-the-art computational epidemic models for HPV disease dynamics, transmission dynamics,  as well as prevention efforts. Selecting 45 most-relevant papers from an initial pool of 10,497 papers identified through keyword search, we classify them based on models used and prevention strategies employed, summarise current research trends,  identify gaps in the present literature, and identify future research directions. In particular, we describe current consensus regarding optimal prevention strategies which favour prioritising teenage girls for vaccination. We also note that optimal prevention strategies depend on the resources available in each country, with hybrid vaccination and screening being the most fruitful for developed countries, and screening-only approaches being most cost effective for low and middle income countries. We also highlight that in future, the use of computational and operations research tools such as game theory and linear programming, coupled with the large scale use of census and geographic information systems data, will greatly aid in the modelling, analysis and prevention of HPV.
\end{abstract}


\section{Introduction}

Human papillomavirus (HPV) is  the pathogen responsible for the most common viral infection of the reproductive system, and is primarily transmitted by sexual contact \cite{Burchell2006EpidemiologyTransmission}. It can be transmitted via bodily fluids, infected genital skin or mucous membranes \cite{WorldHealthOrganization2022HumanPapillomavirus}. Early sexual debut, multiple sexual partners, long-term oral contraceptive uptake, micro element insufficiency, weakened immune system and smoking can aggravate the risk of getting infected by HPV\cite{Rerucha2018CervicalCancer}. About 80\%  of women have been infected by HPV at least once during their lifetime, and 75\% of such infections involve an oncogenic HPV strain \cite{Castellsague2009HPVVaccination}. Most HPV infections can clear spontaneously or become undetectable within two years, if the infected person has a healthy immune system; however, high-risk HPV infections may persist longer and ultimately lead to cervical intraepithelial neoplasia (CIN) \cite{Moscicki2006UpdatingNatural} or cervical cancer.

The link between HPV infection and cervical cancer  was first demonstrated by Schwarz et al.~\cite{Schwarz1985StructureTranscription} by identifying the structure and transcription mechanisms of HPV 16 and 18 pathogens in cervical carcinoma biopsies.  As the topic received greater attention, it was determined that a few of HPV strains were the etiological agents of cervical cancer; as a result, they are now referred  to as the high-risk strains \cite{Villa1997HumanPapillomaviruses}. Among 200 identified HPV types, 12 HPV types (types 16, 18, 31, 33, 35, 39, 45, 51, 52, 56, 58 and 59) are oncogenic and can cause cancer, while  type 68 can also possibly cause cancer \cite{WorldHealthOrganization2022HumanPapillomavirus}.  Oncogenic risk varies by type, whereby  HPV-16 is likely to persist  the longest and also have the  strongest transmissibility among all types \cite{Burchell2006EpidemiologyTransmission, Lowy2008HumanPapillomavirus, WorldHealthOrganization2022HumanPapillomavirus}. HPV-16 is the most prevalent type worldwide, followed by HPV-18 \cite{Lowy2008HumanPapillomavirus, WorldHealthOrganization2022HumanPapillomavirus}. More than 95\% of cervical cancer incidences are caused by persistent oncogenic HPV infection, and among these incidences, the majority are attributable to HPV-16 or HPV-18 strains\cite{WorldHealthOrganization2022HumanPapillomavirus,Carter2011HPVInfection}. 

Cervical cancer is the fourth most diagnosed malignancy among women worldwide, which poses a severe threat to public health. In 2020, over 604,000 women were diagnosed with invasive cervical cancer, and it is estimated that about 341,000 women died of it. The incidence and mortality rates are highest in Eastern Africa, and are almost 10 times as high as those in industrialised regions like Western Asia, Oceania,  and Northern America \cite{Sung2021GlobalCancer}.   In addition, low-risk HPV types such as type 6 and and type 11 are commonly associated with anogenital warts. Therefore, the prevalence of HPV infection puts significant strain on health systems worldwide.

With vaccination and screening, HPV infection and related diseases are preventable \cite{Lowy2008HumanPapillomavirus}. For now, the World Health Organisation (WHO) has approved six commercial HPV vaccinations, including the quadrivalent vaccine firstly licensed in 2006, the bivalent vaccine licensed in 2007 and the nonavalent vaccine licensed in 2014 \cite{WorldHealthOrganization2022CervicalCancer,WorldHealthOrganization2022HumanPapillomavirus}. All vaccines target HPV-16 and HPV-18 strains, and the suggested primary target population is teenage girls aged 9 years or older before their first sexual intercourse (sexual debut) \cite{WorldHealthOrganization2022HumanPapillomavirus}. Following sexual debut, young women are more likely to contract HPV, and the risk increases with new sexual partner acquisition \cite{Burchell2006EpidemiologyTransmission}. In order to intervene before precancerous lesions progress into carcinoma, screening can be used to identify HPV infection and precancerous lesions \cite{Trottier2006EpidemiologyGenital}. WHO suggests that the general population of women start regular cervical screening at the age of 30 years and do it every 5 to 10 years. Women with HIV are advised to receive regular screening every 3 to 5 years starting at age 25 \cite{web2021guideline}. According to empirical studies, HPV screening can effectively reduce incidence of invasive cervical carcinoma \cite{Ronco2014EfficacyHPVbased}.

Epidemic transmission modelling integrates epidemiology with mathematics, computer science, complex systems science and sociology \cite{Duan2015MathematicalComputational}. The multidisciplinary approach serves as a helpful tool to forecast the transmission process  of pathogens and make effective interventions to mitigate outbreaks \cite{Cliff2018InvestigatingSpatiotemporal, zachreson2018urbanization}. Mathematical and computational approaches are the two principal components in epidemic modelling. Mathematical modelling is the typical  initial approach to formulate epidemic transmission, whereby the population is assumed to be homogeneous and well-mixed,  and the transmission process is modelled by a set of variables and simple, often linear relationships between them. The model then develops analytical solutions and  performs theoretical analysis on macroscopic regularities of epidemic spread, thereby predicting  the epidemic threshold and final epidemic magnitude \cite{Duan2015MathematicalComputational, Tuszynski2014MathematicalComputational, Marathe2013ComputationalEpidemiology}. These   assumptions and simplifications enable  mathematical models  to capture concurrent dynamics of outbreaks relatively quickly,  but need to neglect the modelling of attributes and behaviours of  individual people.  Computational modelling  on the other hand focuses on the individual or cohort-level attributes and behaviours,  and  tends to mimic the micro-level dynamics with a higher level of fealty. In the recent decade, with expanding availability of census data and the development of computational power, there has been  a significant increase in the popularity of computational models, which play a considerable role in helping  to understand the multi-level dynamics of epidemics \cite{Duan2015MathematicalComputational, Marathe2013ComputationalEpidemiology}.  Computational models have become increasingly popular as they offer several advantages over purely mathematical models: they are able to  model attributes of individuals with a very high level of detail, they do not require oversimplification of macro-level dynamics, and they have no limitation on the number of dynamics which could be simultaneously modelled, so epidemic spreads in workplaces, schools, households, transportation etc can all be modelled at the same time and their interplay could be understood in great detail.  In the context of HPV infection dynamics, several epidemic transmission models have been developed to predict the final epidemic magnitude and evaluate various intervention strategies \cite{VanDeVelde2010UnderstandingDifferences, Kim2007MultiparameterCalibration, Habbema1985ModelBuilding, Matthijsse2015RoleAcquired, Barnabas2006EpidemiologyHPV, Baussano2013TypeSpecificHuman, Aguilar2015CosteffectivenessAnalysis, Elbasha2008MultiTypeHPV}. These studies vary in terms of the demographics they focus on, the level of detail they capture, the size of population they cover, and the intervention methods they analyse. They also vary in terms of other tools and methods they incorporate, such as computational methods which use network science, agent based models, compartmental modelling, differential equations, and various combinations thereof. Therefore, a rich array of papers which use computational tools to model HPV disease dynamics and intervention methods exist.

In this paper, we conduct a comprehensive review of the computational models that study  HPV  transmission and related prevention strategies, aiming to  investigate and highlight the current trends and future directions of research in eliminating HPV infection using  computational tools and platforms.  Following a well-established  process for selecting papers to review, we select a total of 45 papers from an initial pool of  10,497 candidate papers, and review and analyse them in detail. We show that the existing literature could be meaningfully classified based on  either the type of the computational model used, or the prevention strategy (or strategies) modelled. We identify that there seems to be a consensus in existing literature that females, especially teenage girls, should be prioritised for vaccination. However, we also note that the choice of  effective prevention strategies depends on the resources available to a country, and screening  (used not in conjunction with vaccination) may be the most cost-effective and viable prevention strategy at the moment in low and middle income countries, whereas a hybrid strategy of vaccination followed by screening could produce optimal outcomes in developed countries. We note also that computational tools which are used to model disease dynamics and vaccination uptake in the context of other epidemics, including game theory and linear programming, are not widely used to capture HPV disease dynamics yet, and employing such tools will increase the effectiveness of the computational modelling of HPV. This review is intended to serve as a summary of the state-of-the art and highlight potential future directions in terms of the computational modelling of HPV.

The rest of the paper is organised as follows. In Section \ref{methodology}, we  introduce the taxonomy utilised in this review to classify and categorise the papers that we analyse. In Section \ref{results}, we describe the papers which are covered in this review, summarise their findings, highlight the novelty and importance of each paper, and classify them according to the taxonomy introduced in the previous section. In Section \ref{discussion}, we summarise and comment on current research trends of applying computational models for HPV transmission and prevention, identify gaps in this literature, and based on this propose future research directions. In Section \ref{conclusion}, we provide our conclusions.

\section{Methodology} \label{methodology}

\begin{table}
\tbl{Classification of Epidemic Models}
{\begin{tabular}{m{3cm} m{13cm}}
        \rowcolor{shuangse}
        Type & Methodology \\
        \rowcolor{qianbai}
        & mathematical model, mathematical modelling, male-female mathematical model\\
        \rowcolor{qianbai}
        & deterministic model, heterosexual deterministic model, two-sex deterministic model, deterministic compartmental model, deterministic transmission model, deterministic dynamic model, deterministic dynamic population-based model, deterministic transmission dynamic model, deterministic dynamic compartmental transmission model, decisive model, decision-modelling\\
        \rowcolor{qianbai}
        & dynamic transmission model, dynamic mathematical model, dynamic mathematical transmission model, dynamic compartmental model, transmission dynamic model\\
        \rowcolor{qianbai}
        & compartment model, compartmental model, compartmental deterministic model, compartmental dynamic transmission model, compartmental deterministic dynamic transmission model, open-cohort compartmental model, Kermack McKendrick model, SIR model, SIS model, SIVS model, partial integro-differential equation model, ODE HPV epidemic model, multisite model, CERVIVAC, compartmental dynamic population-based model\\
        \rowcolor{qianbai}
        \multirow{-13}{*}{Mathematical Models} & stochastic model, Markov model, discrete-time Markov approach, Markov cohort model, lifetime Markov cohort model, static Markov model, state-transition Markov model, health state-transition Markov model, individual-based stochastic model, state-transition mathematical model, computer-based Markov model, Excel-bases Markov model, computer-based mathematical model, Markov process model, $n$-state Markov model, life-time multi-stage static Markov approach, static Markov state transition model, Markov state-transition model, Markovian model, $n$-state time-varying transition state model, compartmental Markov model\\
        \rowcolor{xuebai}
        & computational model\\
        \rowcolor{xuebai}
        &system dynamics model, simulation model, macro-simulation model, micro-simulation model, SIVS model, dynamic model, SIRS model, Markov maro-simulation model, stochastic Monte Carlo simulation model, stochastic Monte Carlo computer simulation, transmission dynamic model, dynamic compartment model, semi-Markov model, $n$-state deterministic semi-Markov model, patient level simulation, micro simulation framework, pair model, individual-based stochastic microsimulation model, SIRS-V model\\
        \rowcolor{xuebai}
        &complex network model, random network model, large network model, computational network model\\
        \rowcolor{xuebai}
        \multirow{-8}{*}{Computational Models} &
        agent-based model, stochastic individual-based transmission dynamic model, agent-based simulation model, HPV-ADVISE, STDSIM\\
\end{tabular}}
\label{mathvscomp}
\end{table}

\begin{figure}[ht]
    \centering
    \includegraphics[scale = 0.4]{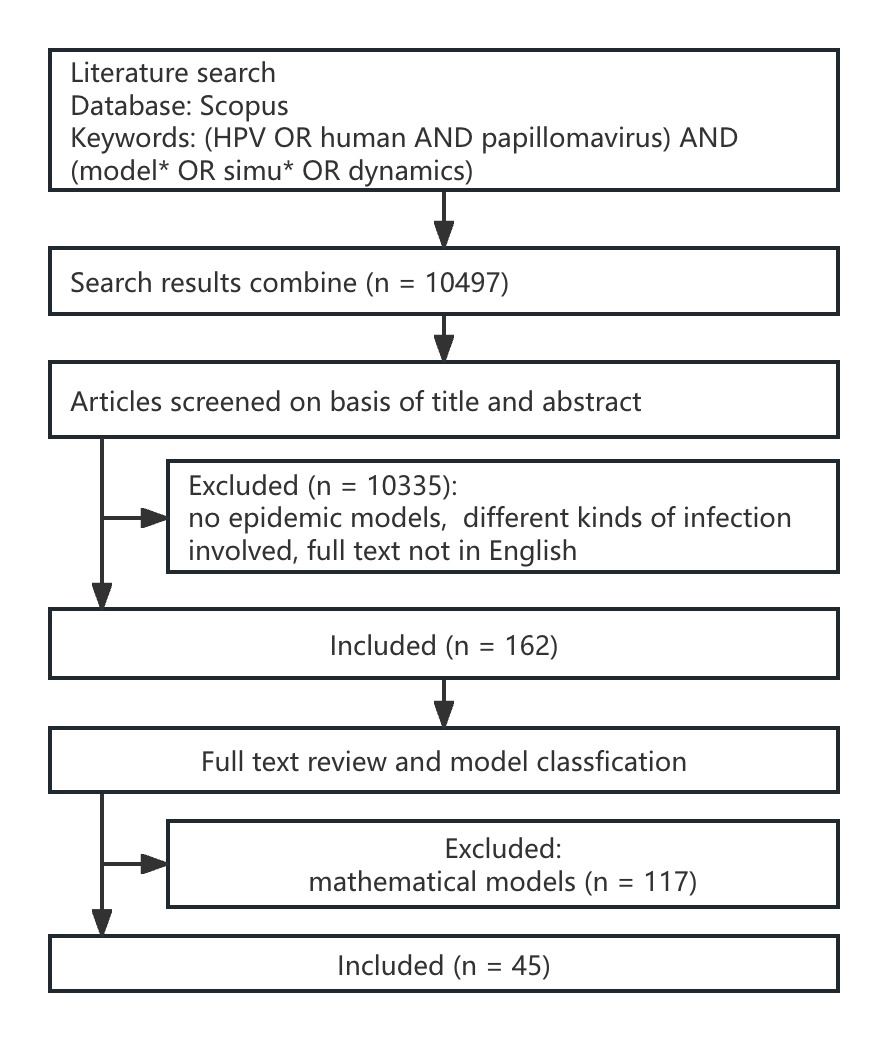}
    \caption{PRISMA Flow Chart. In the beginning, 10497 publications are found in the search results. After the preliminary screening, 10335 publications are screened out, and 162 publications are left. After the secondary screening, 117 publications are excluded. In the end, 15 models and 45 publications are included in the review.}
    \label{prisma}
\end{figure}

A three step process is conducted to select the eligible articles, following the preferred reporting items for systematic reviews and meta-analyses (PRISMA) statement \cite{Liberati2009PRISMAStatement, Chang2020GameTheoretic}.

Firstly, the preliminary selection is carried out in Scopus with keywords including:
\begin{itemize}
    \item  (HPV OR human AND papillomavirus) AND (model* OR simu* OR dynamics)
\end{itemize}

Secondly, highly relevant papers are screened out from the preliminary results by reviewing titles and abstracts according to the following criteria:
\begin{itemize}
    \item \textit{Model:} either a mathematical or computational model should be included in the paper. The models should be based on cohort or individual level with multiple compartments denoting different health statuses; other models, such as cell-based models (e.g. \cite{Asih2016DynamicsHPV}), are not included. The reason why mathematical models are included is that some computational models derived from existing mathematical models; including relevant mathematical models in the process helps with studying structures of computational models in the third step. In addition, mathematical models and computational models may use the same model names, as shown in Table \ref{mathvscomp}, which can be confusing and will need full text review for further classification in the third step.
    \item \textit{Infection:} the modelled infection can only be a specific HPV strain or multiple HPV types. Literature including other infections (e.g. \cite{Tan2018ModelestimatedEffectiveness, Damgacioglu2022LongtermImpact, Schalkwyk2019AreAssociations, VanSchalkwyk2019EstimatedImpact, VanSchalkwyk2021ModellingImpact}) are not considered.
    \item \textit{English:} only records with full texts in English are reviewed; others (e.g. \cite{KlempGjertsen2007CostEffectivenessHuman, DeBlasio2014EstimatingUncertainties}) are not included.
\end{itemize}

Thirdly, full texts of the filtered publications from the last step are reviewed to confirm their eligibility to be included in this review. Literature is further classified according to modelling structure (Section \ref{modelling structure method}). Only computational models are reviewed later in this article.

\begin{itemize}
    \item \textit{Model Structure:} Regardless of model basis, models should include susceptible and infected compartments at least. Those not including HPV infection process (e.g. \cite{Habbema1985ModelBuilding, Blakely2014CosteffectivenessEquity, Mandelblatt2002BenefitsCosts}) are excluded.
    \item \textit{Model Classification:}  Adapted or modified models (e.g. \cite{Miller2015EvaluationNatural, Ryser2015ImpactCoveragedependent, Campos2015WhenHow}) are categorised following the definition of the prime model; if the prime model is defined as a mathematical model by the developers, the adapted or modified models will be classified as a mathematical model (e.g. \cite{Jit2008EconomicEvaluation, Jit2010EstimatingProgression, Choi2010TransmissionDynamic, Chanthavilay2016EconomicEvaluation}). Reimplemented models are considered as independent models.
\end{itemize}

The selection process is shown in Figure \ref{prisma}. Eventually, we have included 15 models and 45 publications in the review. Considering the volume of included literature, we used two criteria to further classify the eligible literature:

\begin{itemize}
    \item \textit{Modelling Structure}
    \item \textit{Prevention Strategy}
\end{itemize}

We  now describe the classification in more detail.

\subsection{Classification based on Modelling Structure} \label{modelling structure method}

\begin{figure}
\centering
\includegraphics[scale = 0.3]{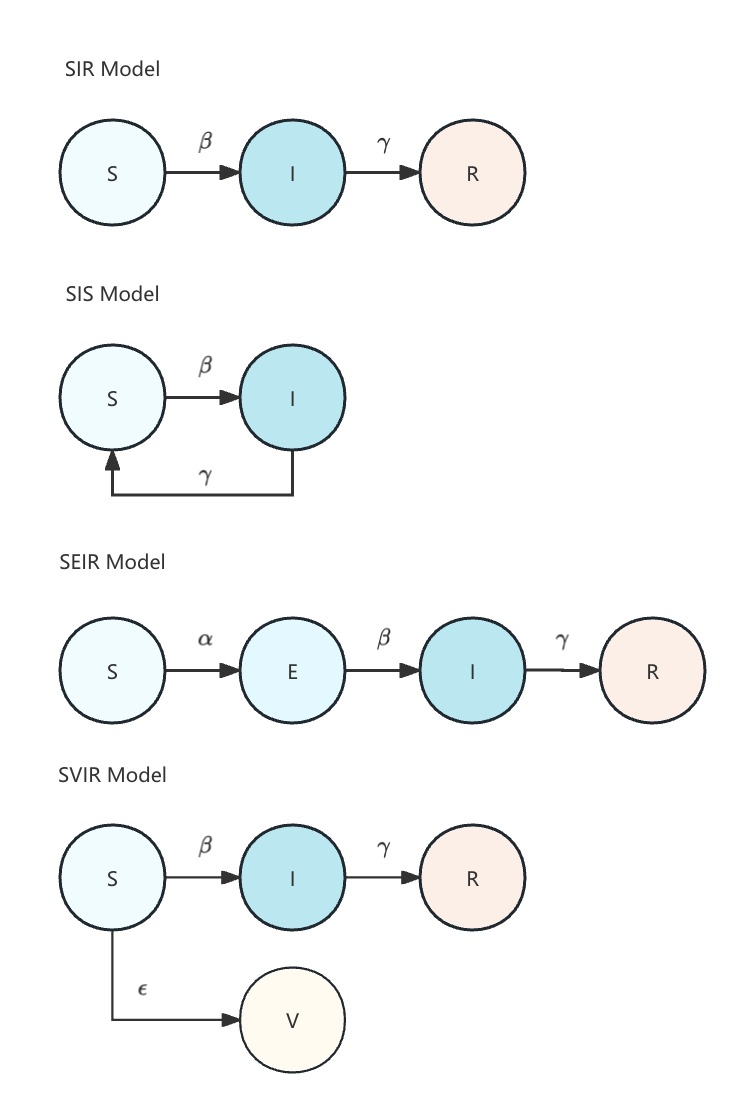}\hspace{5pt}
\caption{In SIR models, susceptible individuals become infected at a rate $\beta$, and infectious individuals can recover and become immune to the infection at a rate $\gamma$. In SIS models, susceptible individuals become infected at a rate $\beta$, and infectious individuals can become susceptible again after recovery at a rate $\gamma$. In SEIR models, susceptible individuals are exposed to infectious individuals at a rate $\alpha$. Exposed individuals contract the infection at a rate $\beta$, and infectious individuals can become resistant to the infection after recovery at a rate $\gamma$. In SVIR models, susceptible individuals are vaccinated and become immune to the infection at a rate $\epsilon$. The rest of them are infected at a rate $\gamma$, and the infectious individuals can recover and also be immune to the infection at a rate $\gamma$.} \label{compartment model fig}
\end{figure}

Computational epidemiology combines mathematical and computational tools to help with understanding the contagion transmission dynamics and developing intervention strategies to contain further dissemination \cite{Patlolla2006AgentbasedSimulation, Marathe2013ComputationalEpidemiology,  Swarup2014ComputationalEpidemiology}. The standard epidemic theory harnesses the concept of compartmentalisation, with the population divided into compartments according to health status and with assumptions about the transfer from one compartment to another \cite{Brauer2008CompartmentalModels, Chang2020GameTheoretic}. 

A classical instance of compartmental models describing the transmission of communicable diseases is the ``S (susceptible), I (infected), R (recovered)'' or SIR model proposed by Kermack and McKendrick in 1927 \cite{Kermack1927ContributionMathematical}, based on comparatively simple assumptions on the rates of flow between different health status groups within the population \cite{Brauer2008CompartmentalModels}. The population $N$ underlying the model is divided into three compartments $S$ (susceptible), $I$ (infective) and $R$ (recovered), and the epidemic process is managed by differential equations:
\begin{align} 
    \frac{dS}{dt} & =-\beta SI 
    \label{d1}\\
    \frac{dI}{dt} & =\beta SI - \gamma I 
    \label{d2}\\
    \frac{dR}{dt} & =\gamma I
    \label{d3}
\end{align}
Here, $\beta$ denotes the transmission rate, and $\gamma$ denotes the recovery rate. When there is no immunity against reinfection, the model is called the SIS (susceptible - infectious - susceptible) model \cite{Brauer2008CompartmentalModels, Chang2020GameTheoretic}. The model structures are shown in Figure \ref{compartment model fig}. The epidemic parameters, $\beta$ and $\gamma$ decide the value of the basic reproduction number of the SIR model:
\begin{equation}
    R_0 = \frac{\beta}{\gamma}
\end{equation}
When the whole population was susceptible at first, $R_0$ denotes the expected number of susceptible individuals that will be infected by one case assuming no individuals can be immunised by vaccination. The basic reproduction number is the threshold quantity determining whether the epidemic will happen or not; the infection will not be able to spread until $R_0 > 1$ \cite{Brauer2008CompartmentalModels, Wang2016StatisticalPhysics}.  

The basic framework of compartmentalisation can be further extended to include more compartments such as E(exposed) for the period, in which susceptible individuals are exposed to infective individuals but are not infectious yet, and V(vaccinated) for the intervention of vaccination \cite{Brauer2008CompartmentalModels}, as shown in Figure \ref{compartment model fig}. The population can incorporate more properties of individuals including age, gender, etc \cite{Chang2020GameTheoretic}.

Computational epidemic modelling approaches can be categorised as follows, according to Luke and Stamatakis's review \cite{Luke2012SystemsScience}:

\begin{itemize}
    \item \textit{System Dynamics Modelling}
    \item \textit{Network-growth based Modelling}
    \item \textit{Agent-based Modelling}
\end{itemize}

\subsubsection{System Dynamics Modelling}

System dynamics models implement simulations as a set of time dependent differential equations that describe interactions between different subpopulations and the transition between compartments \cite{Luke2012SystemsScience}. Within fixed time steps, simulations recursively solve the differential equations and capture the dynamics of epidemic transmission over time \cite{Macal2010AgentbasedSimulation}. For instance, the system dynamics version of SIR model is defined as:
\begin{align} 
    S_{t+1} & =S_t -(\beta S_t I_t /N) \Delta t 
    \label{sd1}
    \\
    I_{t+1} & =I_t +(\beta S_t I_t /N) \Delta t 
    \label{sd2}
    \\
    R_{t+1} & = R_t + (\gamma I_t)\Delta t
    \label{sd3}
\end{align}
The differential equations become:
\begin{align}
    \frac{dS}{dt} & =\frac{\Delta S_{t+1} - \Delta S_{t}}{\Delta t} 
    \\
    \frac{dI}{dt} & =\frac{\Delta I_{t+1} - \Delta I_{t}}{\Delta t} \\
    \frac{dR}{dt} & =\frac{\Delta R_{t+1} - \Delta R_{t}}{\Delta t}
\end{align}
The initial condition is generally defined as:
\begin{equation}
    S_0 = N-1, I_0 = 1, R_0 =0
\end{equation}
Therefore, equations (\ref{sd1})-(\ref{sd3}) have a solution for $\Delta t = 1$, and as $\Delta t \to 0 $, the solution approaches  the solution of equations (\ref{d1})-(\ref{d3}) \cite{Macal2010AgentbasedSimulation}. System dynamics approach strongly emphasises  cohort-based modelling; for more precise simulation results, the population is often stratified into small cohorts according to age, gender, ethnicity, sexual activity, etc. System dynamics models tend to be more flexible to include variables without available empirical data support, and the process of modelling is both iterative and participatory \cite{Luke2012SystemsScience}.

However, a point to be noted that system dynamics models mentioned in empirical publications \cite{Luke2012SystemsScience, Swarup2014ComputationalEpidemiology} only include models implemented with differential equations like the Kermack and McKendrick model introduced above, which are known as deterministic models. In contrast to deterministic approach, stochastic approach introduces randomness to modelling, using transitional probabilities instead of rates to model epidemic process. To a degree, deterministic models can be regarded as an approximation of stochastic models \cite{Allen2008IntroductionStochastic, Chang2020GameTheoretic}. Common stochastic models include discrete time Markov chain (DTMC) models, continuous time Markov chain (CTMC) models, and stochastic differential equation (SDE) models; these approaches varies in terms of the underlying assumptions of time and state variables. In a DTMC model, both time and state variables are discrete. In a CTMC model, state variables remain discrete, while time is continuous. SDE models are based on a transmission process, and both time and state variables are continuous. The principal difference between deterministic models and stochastic models is that even though the deterministic model converges to an endemic equilibrium, the corresponding stochastic model can converge to the disease-free state. The stochastic approaches also have unique properties including the probability of an outbreak, the quasi stationary probability distribution, the final size distribution of an epidemic and the expected duration of an epidemic, which can not be found in deterministic models \cite{Allen2008IntroductionStochastic}.

Here, we assign stochastic models to system dynamics models as a different transition setting approach.

\subsubsection{Network-growth based Modelling}
In epidemics, infectivity is not homogeneous; only a small number of infective individuals known as the super spreaders can transmit the infection to many other susceptible individuals \cite{Brauer2008IntroductionNetworks}. To represent the heterogeneity, Keeling et al.\cite{Keeling1999EffectsLocal, Keeling2005ImplicationsNetwork, Keeling2005NetworksEpidemic} proposed that many infections can be considered to be transmitted through a limited network of contacts with individuals denoted by nodes and contacts between individuals denoted by links. A node in a network is not necessarily an individual; it can also be a cohort or a subpopulation. In unweighted networks, the transmissibility between two nodes is homogeneous; in weighted networks, the transmissibility between two nodes is denoted by the weight of the link. In undirected networks, the infection can be transmitted both ways along the links; in directed networks, the infection is transmitted following the direction of the links. Generally, it is assumed that a node's transmissibility depends on its degree, the number of links it has \cite{Chang2020GameTheoretic}. 

Topology of the underlying network varies according to transmission routes and infection types \cite{Keeling2005NetworksEpidemic}. The networks commonly investigated in the literature of computational epidemiology include lattices, small-world networks, scale-free networks and random networks, etc \cite{Harding2018ThermodynamicEfficiency, Chang2020GameTheoretic}. Of special importance to study epidemic transmission dynamics is scale-free networks. In a network, the degree distribution is denoted as ${p_k}$. In a random network, ${p_k}$ is distributed exponentially with $p_k$ approaching to zero rapidly. While ${p_k}$ of a scale-free network presents a `fatter tail', and the quantity $p_k$  relatively slowly decreases to zero as $k \to \infty$. This corresponds to a situation that there is an active core group and also super spreaders with high degrees in the context of epidemics \cite{Brauer2008IntroductionNetworks}. Scale-free networks exhibit power-law degree distribution whereby the probability mass function (pmf) of degree $P(k)$ is distributed as:
\begin{equation} \label{power law}
    P(k) ~ =  Ak^{-\gamma} 
\end{equation}
where $A$ is a coefficient, and $\gamma$ is known as  the scale-free exponent (also referred as the power-law exponent), with its value typically ranging from 2.0 to 3.0 for real-world scale-free networks \cite{Pastor-Satorras2001EpidemicSpreading, Brauer2008IntroductionNetworks, Bell2017NetworkGrowth, Law2020PlacementMatters}.

The network-growth based approach assigns each node to a neighbourhood that they can only infect
their neighbours and be infected by the neighbours directly \cite{Harding2018ThermodynamicEfficiency}. 
The study of Pastor et al. \cite{Pastor-Satorras2001EpidemicSpreading} exemplified that on a wide range of scale-free networks, the critical threshold of epidemic as $R_0 = 1$ in deterministic models can be absent, and the infection may benefit from the network topology regardless of the transmission rate. Moreover, both a low average degree and a high value of clustering of a network reduce the basic reproduction number $R_0$, comparing to the initial growth rate, which can influence the estimation of epidemic parameters in early stages of the epidemic \cite{Keeling1999EffectsLocal, Keeling2005ImplicationsNetwork}. Therefore, different types of networks can result in varied epidemic behaviours; the underlying topology of a contact network can profoundly affect the infection transmission \cite{Keeling1999EffectsLocal, Keeling2005NetworksEpidemic, Keeling2005ImplicationsNetwork, Brauer2008IntroductionNetworks, Swarup2014ComputationalEpidemiology, Duan2015MathematicalComputational}.

Network-based models have advantages of representing heterogeneously distributed population and interaction patterns among individuals; numerical simulations can help investigate temporal dynamics of epidemic spread \cite{Duan2015MathematicalComputational}. When the contagion is transmitted much faster than the evolution of the contact network, the network can remain static. On the other hand, the network can be adaptive or dynamic to be comparable with the epidemic process. There is increasing attention in amplifying complex network modelling to simulate spatio-temporal epidemic transmission dynamics, and network-based interventions are an active research area in computational epidemiology \cite{Swarup2014ComputationalEpidemiology, Duan2015MathematicalComputational, Chang2020GameTheoretic, Brauer2008IntroductionNetworks}.

In this section, only static networks are considered. Models with underlying static network topology are assigned to network-based models, while the ones built with adaptive or dynamic networks are recognised as agent-based models, which will be introduced next. 

\subsubsection{Agent-based Modelling}

In agent-based models, each individual within the population is modelled as a distinctive agent with attributes and behaviours; interactions between agents provide opportunities to transmit contagion among the population \cite{Bissett2021AgentbasedComputational}. Interdependent agents make interactions to generate ``histories'', and collective properties and behaviours of the overall system will be manifested in simulations \cite{Luke2012SystemsScience, Hedstrom2015RecentTrends}. Agent-based models offer the ability of implementing dynamic interactions between agents, rather than creating static networks, and they can also simulate sophisticated collective phenomena such as neighbourhood effects, markets, social capital, etc \cite{Swarup2014ComputationalEpidemiology}.

Agent-based models represent agents with various attributes and behaviours, a variety of structural and relational constrains and the entwines between local behaviours and global outcomes \cite{Axtell2000WhyAgents}. Characteristics and behaviours of agents can be implemented in programming languages, such as \textit{R}, \textit{Java}, \textit{C++} and \textit{Python}, etc. and also in environments like \textit{NetLogo}\cite{Wilensky2021NetLogo}, \textit{Repast Suite} \cite{CollierRepastSuite}, \textit{AnyLogic} \cite{AnyLogic} etc., which provide visualisation of real-time dynamics. The contact patterns and mobility of agents are determining factors of contagion dissemination. Agents change their location through the simulation and exhibit varied behaviours spatio-temporally, which can shed light on why social distancing is required during a pandemic \cite{Bissett2021AgentbasedComputational}. To better mimic the reality, geographic information systems (GIS) have been utilised to retrieve demographic and environment data to facilitate visualisation of epidemic outbreaks in geographical landscapes \cite{Patlolla2006AgentbasedSimulation, Duan2015MathematicalComputational}. 

Agent-based models have been shown  to be a useful analysis tool in public health research, and are suited in particular to study epidemics dynamics and infectious disease transmission at multiple scales varying from individual communities to the global population\cite{Luke2012SystemsScience}. Their flexibility and generality are their main strengths. Agent-based models hold considerable promise for multi-disciplinary fields, which is not possible for other modelling approaches to handle with similar ease, and the underlying formalism prevents researchers from making assumptions with limited knowledge of the situation \cite{Hedstrom2015RecentTrends}.

Integrating large-scale census datasets can considerably improve the accuracy of epidemiological models; precise implementation of the underlying contact network also plays a crucial role in the simulations \cite{Duan2015MathematicalComputational, Cliff2018InvestigatingSpatiotemporal, Yeung2024AgentBased}. However, the implementation of agent-based models requires data availability, and the simulation need high-performance computational tools. Moreover, with the wide data collection of human behaviour and mobility, the underlying algorithms may be too complex to be formalised \cite{Duan2015MathematicalComputational}.

A key point to be noted here is how we make a deliberate distinction in this review between papers which undertake agent-based modelling,  and papers which undertake  network growth-based modelling, to capture HPV disease dynamics. Indeed, a contact network could be inferred from an agent-based model based on the interactions between individual agents, so it may seem that such an approach uses both agent-based and network-based modelling. However, the primary modelling approach in such cases would be agent-based modelling, and the contact network would be an observable feature of such a model. Such models are classified as `agent-based models' in this review. By contrast, network growth-based models are assumed to be those where the relevant community is primarily captured by a network-growth model, with emphasis on interaction patterns and topological features, rather than individual attributes. In other words, agent-based models are assumed to be those which use high-fidelity data, are computationally intensive, and focus on capturing attributes and transactions of people at individual level (and may be able to indeed express some interaction patterns as networks), while network-growth models are assumed to be those models which do not primarily focus on individual attributes but capture interaction patterns in a topological sense. Therefore, network growth models would typically be less computationally intensive, and they achieve this by capturing interactions only topologically, not spatially.  To stress that the modelled system is created by network growth, and not simply represented as a network, in such models, we call them `network growth-based models' rather than simply `network-based models'.

\subsection{Classification based on Prevention Strategies}

\begin{table}
\tbl{List of HPV Vaccines}
{\begin{tabular}{m{3cm}|m{1.5cm}|m{2.5cm}|m{1.5cm}|m{4cm}}
        \rowcolor{qianbai}
        Vaccine Type & Name & Target Population & Target Age & Routine \\
        \rowcolor{xuebai}
        & & & 9-14 & 2 doses (5-13 months apart) \\
        \rowcolor{xuebai}
        & & & & \\
        \rowcolor{xuebai}
        & \multirow{-3}{*}{\textit{Cervarix}}  & \multirow{-3}{*}{girls and boys} & 15+ & 3 doses (0, 1-2.5 months and 5-12 months)\\
        \cline{2-5}
        
        \rowcolor{xuebai}
        & & & 9-14 & 2 doses (6 months apart) \\
        \rowcolor{xuebai}
        & & & & \\
        \rowcolor{xuebai}
        & \multirow{-3}{*}{\textit{Cecolin}}  & \multirow{-3}{*}{girls} & 15+ & 3 doses (0, 1-2 months and 5-8 months)\\
        \cline{2-5}
        
        \rowcolor{xuebai}
        & & & 9-14 & 2 doses (6 months apart, with a minimum interval of 5 months) \\
        \rowcolor{xuebai}
        & & & & \\
        \rowcolor{xuebai}
        \multirow{-10}{*}{\makecell{bivalent  \\(HPV types 16 and 18)}}& \multirow{-4}{*}{\textit{Walrinvax}}  & \multirow{-4}{*}{girls} & 15+ & 3 doses (0, 2-3 months and 6-7 months)\\

        \rowcolor{yingbai}
        & & & 9-13 & 2 doses (6 months apart) \\
        \rowcolor{yingbai}
        & & & & \\
        \rowcolor{yingbai}
        & \multirow{-3}{*}{\textit{Gardasil}}  & \multirow{-3}{*}{girls and boys} & 14+ & 3 doses (0, 1-2 months and 4-6 months)\\
        \cline{2-5}
        
        \rowcolor{yingbai}
        & & & 9-14 & 2 doses (6 months apart) \\
        \rowcolor{yingbai}
        & & & & \\
        \rowcolor{yingbai}
        \multirow{-6}{*}{\makecell{quadrivalent  \\(HPV types 6, 11 16 \\ and 18)}}& \multirow{-3}{*}{\textit{Cervavax}}  & \multirow{-3}{*}{girls and boys} & 15+ & 3 doses (0, 2 and 6 months)\\

        \rowcolor{yuebai}
        & & & 9-14 & 2 doses (5-13 months apart) \\
        \rowcolor{yuebai}
        & & & & \\
        \rowcolor{yuebai}
        \multirow{-3.5}{*}{\makecell{nonavalent  \\(HPV types 6, 11 16,\\ 18, 31, 33, 45, 52 \\and 58)}}& \multirow{-3}{*}{\textit{Gardasil9}}  & \multirow{-3}{*}{girls and boys} & 14+ & 3 doses (0, 1-2 months and 4-6 months)\\
\end{tabular}}
\label{vaccines table}
\end{table}

For both women and men, HPV infection can result in anogenital warts, respiratory papillomatosis and cancers of the anogenital area, oropharynx, head and neck. For women, persistent oncogenic HPV infection may result in CIN and even progress to invasive cervical cancer (ICC) if not treated. For men, oncogenic HPV infection is associated with a proportion of penile and anal cancer \cite{Lowy2008HumanPapillomavirus, WorldHealthOrganization2022HumanPapillomavirus}.

Natural infection does not induce a vigorous immunological response, and it normally takes 8 to 12 months to develop seroconversion from HPV infection. At present, the available data cannot prove that HPV infection can induce protection against reinfection. Meanwhile, empirical studies suggest that women can be repeatedly infected by the same type or infected by other types concurrently or subsequently \cite{WorldHealthOrganization2022HumanPapillomavirus}. Regarding the relative importance of cervical cancer among HPV-related cancers and the predominant role of HPV infection in cervix, current prevention strategies focus on eliminating cervical cancer but also apply to prevent HPV infection in general and other HPV-related diseases.

\subsubsection{Prevention Approaches} \label{prevention approaches}
HPV infection can be transmitted through sexual contacts; however, traditional approaches of reducing sexually transmitted infections are not typically applicable to HPV infection \cite{Burchell2006EpidemiologyTransmission, Lowy2008HumanPapillomavirus}. Limiting number of sexual partners, practising abstinence and delaying coitarche can reduce the risk but not eliminate it \cite{Burchell2006EpidemiologyTransmission}. HPV infection has high transmissibility and is prevalent in all sorts of sexual contacts \cite{Burchell2006EpidemiologyTransmission, Lowy2008HumanPapillomavirus}. Similarly, circumcision alone cannot prevent HPV infection. Medical data shows high prevalence of HPV infection among circumcised men and their sexual partners, and no evidence has proved that circumcision can reduce the incidence. Condom use is similarly ineffective \cite{Lowy2008HumanPapillomavirus}: even when a condom is used, the exposed part of genital skin may get infected \cite{Burchell2006EpidemiologyTransmission}.

WHO issued a call to eliminate cervical cancer in 2018 and launched the global strategy to accelerate the elimination in 2020. The global strategy includes three complementary pillars: 1) 90\% HPV vaccination coverage of eligible girls; 2) 70\% screening coverage using a high-performance test; and 3) At least 90\% of women with a positive screening test or a cervical lesion to be  treated properly \cite{web2021guideline}.

The recognised best way of preventing cervical cancer is to immunise girls before they are sexually active. Comparing to natural infection, HPV vaccination is more immunogenic and triggers higher polyclonal antibody response, and the avidity of the response will not increase after boosting \cite{WorldHealthOrganization2022HumanPapillomavirus}. HPV vaccines are administrated to prevent cervical premalignant lesions and cancers induced by high-risk HPV infections. The first HPV vaccine was licensed in 2006, and so far, six prophylactic vaccines have been approved. The details of currently available HPV vaccines are listed in Table \ref{vaccines table}. All HPV vaccines contain virus-like particles (VLPs) similar to HPV types 16 and 18, which are the two most prevalent genotypes \cite{Lowy2008HumanPapillomavirus, WorldHealthOrganization2022HumanPapillomavirus}. Empirical experiments suggest that all licensed HPV vaccines perform with  high efficacy in  populations that are not exposed to HPV infections before \cite{WorldHealthOrganization2022HumanPapillomavirus}. Following the onset of sexual activity, adolescents and young women are at the highest risk of acquiring HPV infection, and the risk decreases as age increases \cite{Lowy2008HumanPapillomavirus, Castellsague2009HPVVaccination, Rerucha2018CervicalCancer}. Therefore, HPV vaccination primarily targets  girls aged 9-14 years before their sexual debut. After assuring the high coverage on the primary target population, the secondary target population can include females aged 15+ years, boys in general, and specifically men who have sex with men (MSMs). Till 2022, 125 countries have added HPV vaccination to their national immunisation program for girls, and 47 of them include boys \cite{WorldHealthOrganization2022HumanPapillomavirus}.

The current HPV vaccination cannot provide full protection against all high-risk HPV types, and its impact on women who are vaccinated after the recommended age or not vaccinated is limited. It is argued that,  as a  primary prevention against cervical lesions and cancer, HPV vaccination cannot replace the role of screening later in life \cite{web2021guideline, WorldHealthOrganization2022HumanPapillomavirus}.

To prevent cervical cancer, multiple screening methods are available to identify women who are at risk of, or have cervical precancerous lesions and early invasive cancer \cite{web2021guideline, WorldHealthOrganization2022HumanPapillomavirus}. The conventional approach is the Papanicolaou (Pap) smear test, also known as the cytology test. The cytology test was introduced in the 1940s to detect atypical cells in the cervical epithelium through preparation and interpretation of slides using microscopy. This initiated the transition of switching the emphasis to preventing invasive cancer before it happens \cite{Lowy2008HumanPapillomavirus, web2021guideline, WorldHealthOrganization2022HumanPapillomavirus}. In the last 15 years, newer approaches have been introduced. Visual inspection with acetic acid (VIA) identifies lesions that need treatment or further evaluation on the cervix without magnification. Nucleic acid amplification tests (NAAT) (DNA or mRNA) are performed on molecular level to detect early HPV infection \cite{web2021guideline, WorldHealthOrganization2022HumanPapillomavirus}. Newer techniques and tests have been developed for more precise screening results recently \cite{web2021guideline}. To accelerate the elimination of cervical cancer, WHO strongly recommends that women in general should receive HPV DNA tests as the primary screening test other than cytology or VIA in screening every 5-10 years starting at aged 30 years; for immunocompromised women, HPV DNA tests should be performed every 3-5 years starting at aged 25 years. If the primary screening test result is positive, responsive treatment can be administered based on the test result, which is called the ``screen-and-treat approach''. Otherwise, females with positive results can receive a second test as a ``triage'' test with or without histologically confirmed diagnosis, which is called the ``screen, triage and treat approach'' \cite{WorldHealthOrganization2022HumanPapillomavirus, web2021guideline}. 

In this paper, we classify studies based on prevention approaches as follows:

\begin{itemize}
    \item Screening: screening is the only implemented prevention strategy in the model used by the study.
    \item Vaccination: vaccination is the only implemented prevention strategy in the model used by the study.
    \item Hybrid: both screening and vaccination are implemented in the model used by the study.
\end{itemize}

\section{Results} \label{results}

\begin{figure}[ht]
    \centering
    \includegraphics[width=\textwidth]{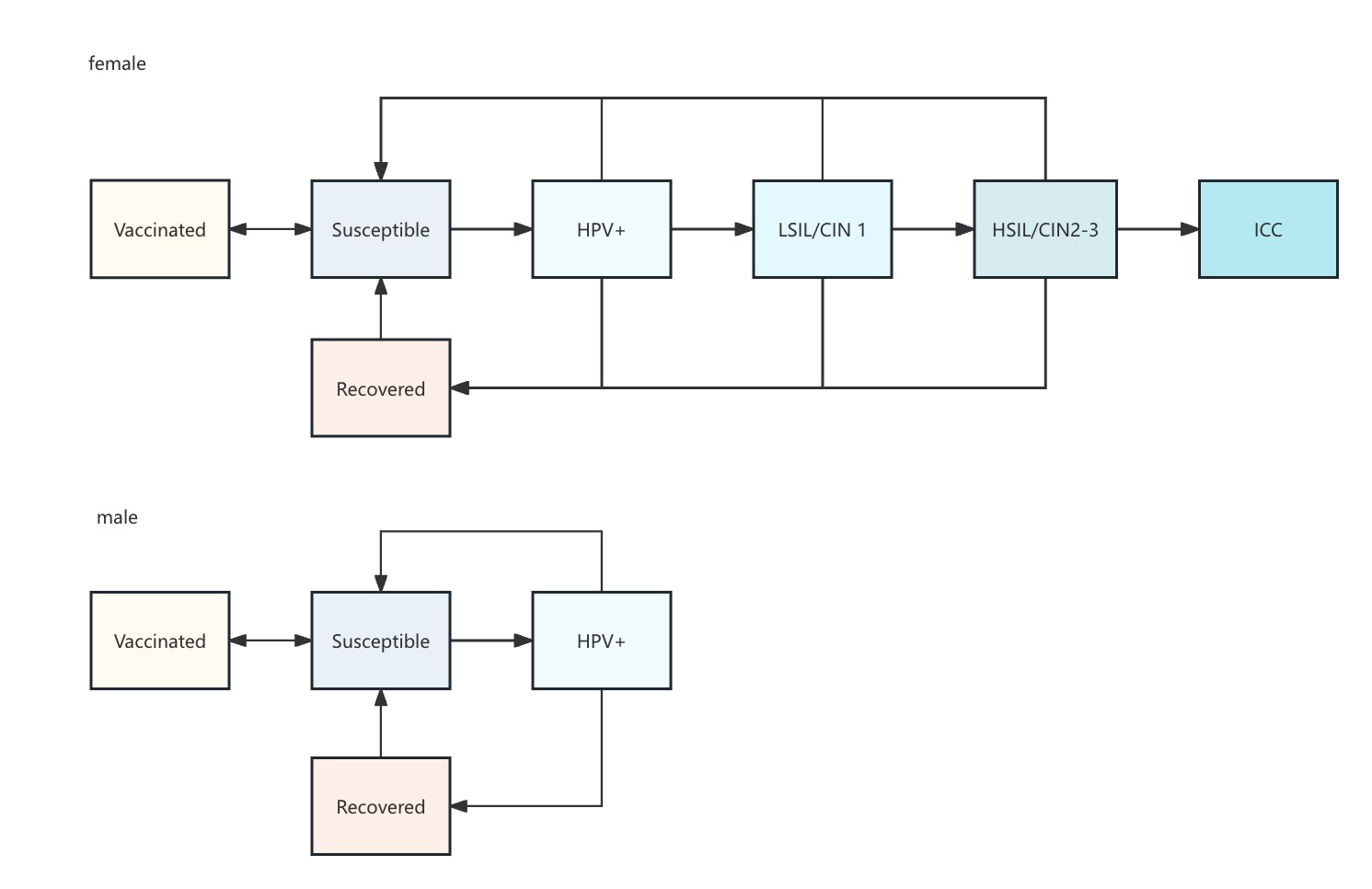}
    \caption{HPV Compartmental models for females and males. Note that the natural progression history of HPV infection differs between females and males. Both vaccinated females and males are immune from HPV infection. Susceptible females contract HPV infection and become infectious. The infectious compartment for females is divided into three stages: 1) an initial stage with positive infection test results; 2) a persistent stage with low-grade squamous intraepithelial lesion (LSIL) or CIN1; 3) a persistent stage with high-grade squamous intraepithelial lesion (HSIL) or CIN2-3, which can progress to invasive cervical cancer (ICC). Before ICC, females in the infectious compartment can progress to the next stage or regress to the previous stage. With viral clearance or responsive treatment, females may be resistant to the infection or become susceptible again. For males, there is only one stage in the infectious compartment. Infected males can acquire immunity against HPV upon viral clearance or responsive treatment, or become susceptible to the infection again. }
    \label{HPV compartment model fig}
\end{figure}

\subsection{Classification based on Modelling Structure} 
Mainly primary models are reviewed in this section,  organised based on the models  used in them. Adapted models are considered to maintain modelling structures from the primary models. If there is major modification in an adapted model, it will have been considered as a reimplemented model in the third step of the PRISMA review in Section \ref{methodology}. 

It should be noted that, considering the natural progression history of HPV infection, besides the basic compartments including susceptible, infectious, recovered and vaccinated introduced in Section \ref{modelling structure method}, the infectious compartment can be further stratified into sub-compartments for females in the model. Due to the difference between female and male body structure, cervical abnormalities and cervical cancer can only be found among females \cite{Moscicki2006UpdatingNatural}; therefore, in natural history of HPV infection and progression into squamous cell carcinoma (SCC) and cervical cancer models, female individuals are assigned more health states, such as different levels of CINs and cervical cancer. A possible compartmental framework  taking this into account is shown in Figure \ref{HPV compartment model fig}.

\subsubsection{System Dynamics Models}

\begin{table}
\tbl{Classification of System Dynamics Models}
{\begin{tabular}{m{2cm} m{2cm} m{4.5cm} m{4.5cm}}
        \rowcolor{qianbai}
        Model Basis & Transition Setting & Primary Models & Adapted Models \\
        \rowcolor{xuebai}
        \cellcolor{shuangse}&& Al-arydah and Smith 2011 \cite{Al-arydah2011AgestructuredModel} & Al-arydah and Malik 2017 \cite{al2017age}
        \\
        \rowcolor{xuebai}
        \cellcolor{shuangse}&& Jarynowski and Serafimovic 2014 \cite{Jarynowski2014StudyingPossible} &
        \\
        \rowcolor{xuebai}
        \cellcolor{shuangse}&&Madhu and Al-arydah 2021 \cite{Madhu2021OptimalVaccine}& 
        \\
        \rowcolor{xuebai}
        \cellcolor{shuangse}&& Berhe and Al-arydah 2021 \cite{Berhe2021ComputationalModeling} & 
        \\
        \rowcolor{xuebai}
        \cellcolor{shuangse}&\multirow{-6}{*}{deterministic}&Al-arydah 2021 \cite{al2021two}   & Al-arydah 2023 \cite{Al-arydah2023MathematicalModeling}
        \\

        \rowcolor{yingbai}
        \cellcolor{shuangse}&& Burchell et al. 2006 \cite{Burchell2006ModelingSexual} &\\
        \rowcolor{yingbai}
        \cellcolor{shuangse}&& Van de Velde et al. 2007 \cite{VanDeVelde2007ModelingHuman} &\\
        \rowcolor{yingbai}
        \cellcolor{shuangse}&& Coupé et al. 2009 \cite{Coupe2009HPV1618} & Coupé et al. 2009 \cite{Coupe2009HowScreen} \\
        \rowcolor{yingbai}  
        \cellcolor{shuangse} && Bogaards et al. 2010 \cite{Bogaards2010ModelbasedEstimation} & Bogaards et al. 2011 \cite{Bogaards2011LongtermImpact} \newline Bogaards et al. 2011 \cite{Bogaards2011SexspecificImmunization}
        \\
        \rowcolor{yingbai}
        \cellcolor{shuangse}\multirow{-12}{*}{cohort}&\multirow{-6}{*}{stochastic}& Landy et al. 2018 \cite{Landy2018WhatCervical} &
        \\

        \rowcolor{baifen}
        \cellcolor{xiangya}&& Kim et al. 2007 \cite{Kim2007MultiparameterCalibration} &Goldie et al. 2007 \cite{Goldie2007CosteffectivenessHPV}\newline Goldhaber-Fiebert et al. 2007 \cite{Goldhaber-Fiebert2007ModelingHuman} \newline Goldhaber-Fiebert et al. 2008 \cite{Goldhaber-Fiebert2008CosteffectivenessCervical} \newline Kim et al. 2008\cite{Kim2008ExploringCosteffectiveness}\newline Diaz et al. \cite{Diaz2008HealthEconomic}\newline Kim and Goldie et al. 2008\cite{Kim2008HealthEconomic}\newline Kim and Goldie et al. 2009 \cite{Kim2009CostEffectiveness}\newline Burger et al. 2012 \cite{Burger2012CosteffectivenessCervical}\newline Campos et al. 2012 \cite{Campos2012HealthEconomic}\newline Campos et al. 2014 \cite{Campos2014UpdatedNatural}\newline Campos et al. 2015 \cite{Campos2015ComparativeCosteffectiveness}\newline Campos et al. 2015 \cite{Campos2015WhenHow}\newline Campos et al. 2017 \cite{Campos2017EstimatingValue}\\

        \rowcolor{baifen}
        \cellcolor{xiangya}&& Coupé et al. 2012 \cite{Coupe2012ImpactVaccine} &\\
        \rowcolor{baifen}        \cellcolor{xiangya}&&Ekwunife and Lhachimi 2017 \cite{Ekwunife2017CosteffectivenessHuman}&
        \\
        \rowcolor{baifen}
        \cellcolor{xiangya}\multirow{-16}{*}{individual}&\multirow{-16}{*}{stochastic}& Datta et al. 2019 \cite{Datta2019AssessingCosteffectiveness} & \\
\end{tabular}}
\label{system dynamics models}
\end{table}

System dynamics models inherit the frameworks of mathematical models and incorporate them with computer simulations \cite{Patlolla2006AgentbasedSimulation, Swarup2014ComputationalEpidemiology}. System dynamics models are developed following the compartmentalisation concept that divides the population into exclusive compartments of different health statuses, as mathematical models do. The transitions that individuals move from one compartment to the next are captured as the epidemic dynamics \cite{Brauer2008CompartmentalModels, Chang2020GameTheoretic}. Generally, all individuals are susceptible to the infection in the beginning. When the infection seed is introduced to the population, the epidemic process starts. Susceptible individuals encounter infectious individuals, which enables the dissemination of the infection; susceptible individuals will be moved to the infectious compartment if they are infected during the encounter. The infectious individuals will be immune to the infection after they recover; these individuals will be moved from the infectious compartment to the recovered compartment. When there is no more individuals that can be infected, the epidemic will gradually die out. This is the basic susceptible(S)-infectious(I)-recovered(R) compartmental framework \cite{Brauer2008CompartmentalModels, Chang2020GameTheoretic}. If no long-term immunity can be acquired from the infection, infected individuals will be moved to the susceptible compartment after they recover, which is the SIS approach. In addition, if the encounter is supposed to happen at a rate, an E(exposed) compartment will be added between S and I compartments. The epidemic dynamics are similar to what is shown in Fig \ref{compartment model fig}. Vaccinated individuals will be marked as an individual V(vaccinated) compartment in the process. More compartments can be added if needed \cite{Brauer2008CompartmentalModels}. 

Due to the relatively high volume of system dynamics models included in the literature, the models are further classified according to the following criteria:

\begin{itemize}
    \item \textit{Model Basis}
    \item \textit{Transition Setting}
\end{itemize}

According to model basis, system dynamics models can be divided into cohort based models and individual based models. If the epidemic process is defined by a set of differential equations, for instance, as shown in Eq. [\ref{d1}]-[\ref{d3}], where the transitions are decided by constant rates, the model is deterministic. In other words, all individuals in the same compartment share the same transitional rates. For stochastic models, the differential equations are replaced by transitional probabilities with the introduction of randomness. Based on the criteria, the system dynamics models are divided into cohort deterministic models\cite{ Jarynowski2014StudyingPossible, Berhe2021ComputationalModeling}, cohort stochastic models\cite{Burchell2006ModelingSexual, VanDeVelde2007ModelingHuman, Coupe2009HPV1618, Bogaards2010ModelbasedEstimation, Landy2018WhatCervical}, and individual stochastic models\cite{Kim2007MultiparameterCalibration, Coupe2012ImpactVaccine, Ekwunife2017CosteffectivenessHuman, Datta2019AssessingCosteffectiveness}. This is summarised in Table \ref{system dynamics models}.

In cohort-based deterministic models \cite{Berhe2021ComputationalModeling, Jarynowski2014StudyingPossible}, the population is stratified into cohorts based on gender, age, etc. The progression and regression rates stay constant, and the epidemic process is modelled by differential equations. Berhe and Al-arydah \cite{Berhe2021ComputationalModeling} developed a heterosexual deterministic SIVS HPV model. The female population was divided into susceptible unvaccinated females, infected females and susceptible vaccinated females, and the  male population was compartmentalised similarly. Susceptible unvaccinated females and males received impulsive vaccination periodically, and a certain proportion of susceptible vaccinated individuals might lose their immunity from the vaccination and become susceptible unvaccinated each year. During the period between two consecutive pulse shots, the model worked as a standard SIVS model. In a similar work, Jarynowski and Serafimovic \cite{Jarynowski2014StudyingPossible} extended the classical SIR model. The male population was divided into Susceptible (Sm), Infectious (HPVm) and Recovered (Rm). The female population included ``Long-term colonised (StageIIw)'', ``Vaccinated (Vw)'' and ``Having cancer (Cancer)'', besides Sw, HPVw and Rw. All relationships shared the same infectivity, and partner change rates were stochastic according to sexual active levels and age. Progression and regression rates between compartments were implemented to be age-specific, and the epidemic process was again modelled by differential equations.

Cohort-based models \cite{Burchell2006ModelingSexual, VanDeVelde2007ModelingHuman, Coupe2009HPV1618, Bogaards2010ModelbasedEstimation, Landy2018WhatCervical} also utilised transitional probabilities from stochastic models. Burchell et al. \cite{Burchell2006ModelingSexual} employed a heterosexual stochastic Monte Carlo simulation approach to implement sexual contact dynamics. Accordingly, each cohort shared the same intercourse frequency, and condom use frequency was assigned to each couple. HPV transmissibility was set as male-to-female and designated per coital act in 12 different levels. 27 HPV types were considered in the model. Van de Velde et al. \cite{VanDeVelde2007ModelingHuman} modelled four HPV genotypes: HPV-16, HPV-18, other high-oncogenic risk types and low-risk oncogenic types and incorporated dual between the four genotypes. The model implemented the natural history of HPV infection and SCC in a woman cohort during their lifetimes with six mutually exclusive compartments: susceptible, immune, infected, CIN1, CIN2/3, and SCC.  In the model, individuals could develop lifelong immunity after infection or CIN clearance, or become susceptible again. The transitional probabilities were designed to vary depending on age and HPV genotype. It should be noted that here the model included a separate compartment, which derived from CIN2/3, representing individuals having hysterectomy. This cohort would not become susceptible or resistant to the infection. Coupé et al. \cite{Coupe2009HPV1618} also followed health trajectories of women health over their lifetimes. They implemented 14 HPV infections in parallel and allowed concurrent infections. Progression probabilities were type-specific. In the model, females could develop CIN1 without HPV infection, and viral clearance and lesion regression were possible before CIN3. Progression probabilities were stochastic according to HPV genotype, age, and compartments. No immunity was implemented since this is an SIS model. Bogaards et al. \cite{Bogaards2010ModelbasedEstimation} developed a dynamic model to mimic heterosexual partnership formation and also implemented 14 high-risk HPV infection. Only women were included in the SIRS model, whose structure is similar to the model developed by Van de Velde et al. \cite{VanDeVelde2007ModelingHuman} without the hysterectomy compartment. Though the model incorporated relationship formation, transmissibility was independent of sexual activity and infection stage. Landy et al. \cite{Landy2018WhatCervical} ran HPV16/18 and non-16/18 high-risk HPV transmission independently among women aged 12-80 in an SIS model. HPV infections were classified into new HPV infection and persistent HPV infection with the same parameter for progression, and high-grade CIN would progress into asymptomatic cancer first then symptomatic cancer. The transitional probabilities were specific to age and type, and before asymptomatic cancer, infection and CIN could possibly clear spontaneously.

Similar to cohort-based stochastic models, individual-based stochastic models replace certain transitional rates with heterogeneous probabilities. Kim et al. \cite{Kim2007MultiparameterCalibration} and Coupé et al. \cite{Coupe2012ImpactVaccine} model women  from prior to their sexual debut till the very end of their life and include compartments of HPV infection and natural history of cervical cancer. Kim et al.\cite{Kim2007MultiparameterCalibration} simulated four HPV strains: low-risk HPV, HPV-16, HPV-18 and other high-risk HPV; Coupé et al. \cite{Coupe2012ImpactVaccine} employed a Markov chain that runs independently for 14 high-risk HPV genotypes. Both models implemented transmission probability and progression rates as type dependent and regression rates as homogeneous. Similarly, Kim et al.\cite{Kim2007MultiparameterCalibration} included different stages of cervical cancer: local cancer, regional cancer and distant cancer, which can not spontaneously clear or be cured in the model. Ekwunife and Lhachimi \cite{Ekwunife2017CosteffectivenessHuman} remodelled a patient level simulation of natural history for cervical cancer disease from an empirical model-based study \cite{Demarteau2014ModelingOptimal}. The simulation commenced when women were 9 years old and terminated when they are 99 years old. The model unified 12 oncogenic HPV genotypes as a single HPV infection and signified nine mutually exclusive heath states. Here, before cancer state, all states could lead to mortality in general, and cervical cancer can be cured or lead to death due to cancer or in general. Transitional probabilities included both certain rates and stochastic probabilities depending on age. Datta et al.\cite{Datta2019AssessingCosteffectiveness} developed a heterosexual SIRS-V model with compartments including susceptible, infected, recovered and vaccinated. Natural immunity acquired from HPV infection is defined here as having a  short-duration; only vaccination can provide longer-lasting protection. The population is divided into cohorts according to age, sex, sexual orientation and previous sexual experience. Partnership formation is implemented, and nine independent HPV genotype transmissions are modelled independently. Transmission probability and recovery rate differ between females and males. The asymmetric transmissibility between partners depend on individual characteristics and HPV types, and the recovery rate only depends on sex.

\subsubsection{Network-growth based Models}

\begin{table}
\tbl{Network growth-based and Agent-based HPV Models}
{\begin{tabular}{m{4cm} m{5cm} m{6cm}}
        \rowcolor{qianbai}
        Modelling Structure & Primary Model & Adapted Models \\
        \rowcolor{shuangse}
        \cellcolor{xuebai}&& Acedo et al. 2018 \cite{Acedo2018CalibratingLarge}\\
        \rowcolor{shuangse}
        \cellcolor{xuebai}&& Villanueva et al. 2019\cite{Villanueva2019CalibrationAgentbased} \\
        \rowcolor{shuangse}
        \cellcolor{xuebai}&& Muño-Quiles et al. 2021 \cite{Munoz-Quiles2021EliminationInfections} \\
        \rowcolor{shuangse}
        \cellcolor{xuebai}\multirow{-4}{*}{Network-growth-based Models}&\multirow{-4}{*}{Díez-Domingo et al. 2017 \cite{Diez-Domingo2017RandomNetwork}} & Villanueva et al. 2022 \cite{Villanueva2022MathematicalModel} \\
        \rowcolor{xiangya}
        \cellcolor{yudu}&Olsen and Jepsen 2010 \cite{Olsen2010HumanPapillomavirus} &\\
        \rowcolor{baifen}
        \cellcolor{yudu}& & Brisson et al. 2011 \cite{Brisson2011IncrementalImpact}\\
        \rowcolor{baifen}
        \cellcolor{yudu}& &  Van de Velde et al. 2012 \cite{VanDeVelde2012PopulationlevelImpact}\\
        \rowcolor{baifen}
        \cellcolor{yudu}& & Malagón et al. 2013 \cite{Malagon2013ImpactDifferential}\\
        \rowcolor{baifen}
        \cellcolor{yudu}& & Laprise et al. 2014 \cite{Laprise2014ComparingCosteffectiveness}\\
        \rowcolor{baifen}
        \cellcolor{yudu}& \multirow{-5}{*}{Van de Velde et al. 2010 \cite{VanDeVelde2010UnderstandingDifferences}} & Drolet et al. 2017 \cite{Drolet2017ImpactHuman} \\
        \rowcolor{gao}
        \cellcolor{yudu} \multirow{-7}{*}{Agent-based Models}&Matthijsse et al. 2015 \cite{Matthijsse2015RoleAcquired} &\\
\end{tabular}
\label{network and abm models}}
\end{table}

Network-based approach extends system dynamics to graphs to emphasise the relationships among sets of actors \cite{Swarup2014ComputationalEpidemiology, Luke2012SystemsScience}. Network analysis has been widely utilised in public health especially disease transmission in this century \cite{Luke2012SystemsScience}. Representing interpersonal relationships with abstract network can help analyse the underlying topology and understand individual behaviour and adaption, which give insights into contagion transmission and prevention strategies \cite{Marathe2013ComputationalEpidemiology}. Back in the early years of the HIV epidemic, epidemiologists incorporated the fundamental SIR model with the local social network information to visualise the disease transmission and investigate feasible interventions \cite{Luke2012SystemsScience}.

In network-based models, individuals or subpopulations are denoted by nodes, and the interplay between them are represented by links. Generally, the transmissibility of a node depends on the degree of the node, which is the number of links it has \cite{Chang2020GameTheoretic}. Two networks with the same size  can display various structures, and different topologies can represent varied transmission routes and infection types \cite{Keeling2005NetworksEpidemic}. The networks commonly investigated in the literature of computational epidemiology include lattices, small-world networks, scale-free networks and random networks \cite{Harding2018ThermodynamicEfficiency, Chang2020GameTheoretic}. Empirical studies demonstrated that the underlying topology of a contact network can impact the infection transmission over the network \cite{Keeling1999EffectsLocal, Keeling2005ImplicationsNetwork, Keeling2005NetworksEpidemic, Brauer2008IntroductionNetworks, Duan2015MathematicalComputational}. For instance, the epidemic threshold $R_0$ approaches zero in a scale-free network  of infinite size so that large outbreaks can emerge with fairly low transmissibility \cite{Swarup2014ComputationalEpidemiology}.

In the existing literature that we review, there is only one model \cite{Diez-Domingo2017RandomNetwork} implemented which has an underlying static network; to better introduce the model, variations of this \cite{Acedo2018CalibratingLarge, Villanueva2019CalibrationAgentbased, Munoz-Quiles2021EliminationInfections, Villanueva2022MathematicalModel} are included, as shown in Table \ref{network and abm models}. Díez-Domingo et al. \cite{Diez-Domingo2017RandomNetwork} constructed a random network model to mimic sexual contacts. The original network was designed as a heterosexual network, then nodes representing men who have sex with men (MSM), who might also have female sexual partners, were later added into the model \cite{Munoz-Quiles2021EliminationInfections, Villanueva2019CalibrationAgentbased, Villanueva2022MathematicalModel},  In the sexual contact network, each node denoted an individual, and basic characteristics including age and gender were randomly assigned according to population histogram. The total number of lifetime sexual partners (LSP) representing sexual behaviours was distributed to each node proportionally considering assortativity, which is the attribute quantifying the tendency of nodes connecting with similar  nodes in networks \cite{Piraveenan2007InformationCloningScaleFree, Piraveenan2009LocalAssortativity, Lizier2011FunctionalStructural, Law2020PlacementMatters, chang2020impact}. Most individuals had four or less LSPs. Formation of relationships depended on age and LSP for both partners, and the weight function pairing nodes $i$ and $j$ was defined as follows:
\begin{equation}
    \pi(i, j) =
    \begin{Bmatrix}
        |nLSP[i] - nLSP[j]|  &  nLSP[i], nLSP[j] \leq 4 \\
        0  & nLSP[i], nLSP[j] > 4 \\
        100 & otherwise
    \end{Bmatrix}
    + |Age[i]-Age[j]-1.8|
\end{equation}
where $nLSP[i]$ denoted the expected LSP of node $i$. $Age[i]$ was the age of node $i$, and $1.8$ was the average age difference of couples in Spain back then. To form an assortative network, the lower the weight was, the more likely the relationship would form. Nodes with similar age and similar behaviours represented by LSP had more chances to pair with each other; couples where one owned few LSP and the other had way more LSP would be uncommon \cite{Diez-Domingo2017RandomNetwork, Villanueva2022MathematicalModel}. During the network formation, women were prioritised to be paired in descending order of LSP in order to optimise the pairing process, and each MSM was paired with a woman to avoid isolated network components \cite{Acedo2018CalibratingLarge}. In the network, sexual contact frequency was age-dependent \cite{Diez-Domingo2017RandomNetwork}. 

The primary model \cite{Diez-Domingo2017RandomNetwork} simulated two HPV strains: high-risk HPV and low-risk HPV, and later the infection types were stratified to specific HPV genotypes \cite{Villanueva2022MathematicalModel}.
Transmission probability depended on the age groups where the two individuals belong to, infection type and transmission route. Men were estimated to have higher probabilities to transmit contagion to women. Here, randomness was introduced to the model. Certain probabilities were set as epidemic parameters over the network. In order to constrain that a contagion was transmitted by sexual contacts, random numbers were generated to see if they were under the given thresholds decided by the epidemic parameters. If the random numbers met the requirements, the transmission was successful. When a susceptible individual was infected during sexual contacts, the infection duration (clearance time) would be randomly assigned from the expected intervals \cite{Villanueva2019CalibrationAgentbased}. No natural immunity or reinfection was mentioned in the model.

\subsubsection{Agent-based Models}
Agent-based models use bottom-up implementation to simulate epidemic transmission dynamics by examining how individual's attributes, their interactions and the environment result in collective emergent behaviour \cite{Luke2012SystemsScience, Duan2015MathematicalComputational}. In agent-based models, an agent represents a distinctive individual with traits and behaviours, and the agents interact with each other, which provides opportunities to transmit contagion among the population \cite{Bissett2021AgentbasedComputational}. Similar to network-based models, agent-based models also utilise underlying network infrastructure to represent connections among the agents; however, to demonstrate the interaction dynamics, agent-based models construct adaptive or dynamic networks as opposed to static networks in network-based models \cite{Swarup2014ComputationalEpidemiology}. In the dynamic system, the local contact patterns and mobility change consecutively as the agents' behaviours vary throughout the simulation, which further impact the global contagion transmission \cite{Axtell2000WhyAgents, Bissett2021AgentbasedComputational}. In agent-based models, there is no formalised definition of the global system behaviour like system dynamics models, which result in a form of decentralised modelling \cite{Luke2012SystemsScience}. To understand the collective properties and behaviours of the global system, local interactions are captured during simulations \cite{Luke2012SystemsScience, Hedstrom2015RecentTrends}. Agent-based models also offer the benefit of modifying agents' attributes, behaviours, interactions and epidemic parameters to assess their impact on producing the population-level outcome and generate more complex phenomena such as social capital, neighbourhood effects, etc. \cite{Swarup2014ComputationalEpidemiology, Bissett2021AgentbasedComputational}. With the advantage of solving problems with a multi-disciplinary approach and providing realistic approximations, agent-based models are becoming increasingly popular in computational epidemiology \cite{Hedstrom2015RecentTrends, Bissett2021AgentbasedComputational}.

In the existing literature that we  review, three agent-based models \cite{Olsen2010HumanPapillomavirus, VanDeVelde2010UnderstandingDifferences, Matthijsse2015RoleAcquired} are proposed, as shown in Table \ref{network and abm models}.

Olsen and Jepsen \cite{Olsen2010HumanPapillomavirus} constructed a closed ``mini-society'' of 25,000 individuals that constituted a heterosexual contact network in the simulation. The agents were aged 10-78 years, and the distribution mimicked the general population in Denmark back then. The gender ratio was approximately 50\%. Here, the agents' interactions are implemented by if/then logic depending on agents' attributes instead of equations, and the sexual act frequency is assigned by a random-gamma distribution with a mean of 9.48 per month. It should be noted that this model allowed concurrent partners, and each relationship was assigned a duration from a normal distribution based on age in months, as follows:
\begin{equation}
    Y \sim |\mathcal{N} (0.8 \times {age} - {12}, {age} / {0.5}) \times 12|
\end{equation}
where the older the agents were, the longer the relationship maintained. Four HPV genotypes, HPV-6/11/16/18 corresponding to quadrivalent HPV vaccination, were allocated to each age group according to the prevalence in the Danish population. The risk and duration of infection were implemented type-specific and stochastic. As low-risk HPV strains, HPV types 6 and 11 were implemented with a different progression history from the high-risk strains; agents with HPV-6/11 infection could only progress to genital warts instead of cervical lesions. Female agents with high-risk HPV infection could progress to CINs and further cervical cancer. No natural immunity is implemented in the model; therefore, after viral clearance, the risk of being reinfected by the same infection would not change. During the health trajectories, the progression and regression probabilities were defined by estimated parameters in years. 

Van de Velde et al. \cite{VanDeVelde2010UnderstandingDifferences} developed a stochastic individual-based HPV transmission model with sequential partnership formation and dissolution, which was later named as HPV-ADVISE \cite{VanDeVelde2012PopulationlevelImpact, Malagon2013ImpactDifferential, Laprise2014ComparingCosteffectiveness, Drolet2017ImpactHuman}. Individuals joined the simulated population prior to sex debut, and they were assigned characteristics including age, gender, sexual activity level, onset of sexual activity, current health status and current partnership status upon entering the simulation \cite{VanDeVelde2010UnderstandingDifferences, Brisson2011IncrementalImpact, VanDeVelde2012PopulationlevelImpact,
Malagon2013ImpactDifferential, Laprise2014ComparingCosteffectiveness, Drolet2017ImpactHuman}. Without the HPV infection, the model was referred as the demographic model. In the demographic model, the population stayed open and stable, in which individuals enter and die following the mortality rate specific to age and gender to mimic the national demography \cite{VanDeVelde2010UnderstandingDifferences}. During the simulation, potential sexual behaviour, health-seeking behaviour (vaccination and screening), etc. are recorded for each agent \cite{VanDeVelde2012PopulationlevelImpact}. The underlying sexual contact pattern was represented by a stochastic pair formation and separation process. Sequential monogamous stable and instantaneous relationships were implemented in the model; moreover, this model also considered about casual sexual relationships between men and female sexual workers (FSW). Partnership formation and termination were initiated by female agents; the rates of new partner acquisition and partnership termination was related to females' age, sexual activity level and current partnership status \cite{VanDeVelde2010UnderstandingDifferences}. Single females formed a new partnership at a rate, which could become stable with a probability. To form the partnership, age and sexual activity level on both sides would be considered, which implied assortativity in the sexual contact pattern; the type of partnership is based on the proportion of female agents' having sex before marriage. In a stable relationship, the rate of separation, total number of sexual contacts and partnership duration solely depended on the female's age and sexual activity level. For instantaneous partnerships, a small value is assigned as the total number of sexual acts with a fairly short duration \cite{VanDeVelde2010UnderstandingDifferences, VanDeVelde2012PopulationlevelImpact}. The model implemented transmission and natural progression history of 18 HPV types including both low-risk and high-risk strains, and each type was modelled individually without co-infections; therefore, an individual can be infected by multiple genotypes simultaneously \cite{Brisson2011IncrementalImpact}. In the natural history of HPV infection, infected females could clear the infection and become immune or susceptible, or remain infected or progress to more severe stages; females with CIN might regress to a less severe stage or even fully recover to immune or susceptible status \cite{VanDeVelde2010UnderstandingDifferences}. The infection transmission was implemented per sex act, and the model allowed male-to-female and female-to-male transmission probabilities to be different. Since the transmission probability per sex act was gender and type-specific, and the frequency of sex acts was decided by age and level of sexual activity of the female; therefore, the transmission probability per relationship was age, gender and level of sexual activity-specific. Infection clearance rates were related to infection type, patient's age and gender. Only females could develop life-long immunity from infection by stochastic processes \cite{VanDeVelde2010UnderstandingDifferences}. To be noted, though 18 HPV types were represented in the model, quadrivalent vaccine was implemented against HPV-16, 18, 6 and 11. Three characteristics of the vaccine were modelled: 1) vaccine uptake probability; 2) vaccine efficacy per sex act; and 3) duration of protection sampled from a normal distribution. The baseline scenario was the vaccination coverage reached 100\% with 95\% efficacy, and the duration was either lifelong or averaged 20 years \cite{VanDeVelde2010UnderstandingDifferences, VanDeVelde2012PopulationlevelImpact}.

Matthijsse et al. \cite{Matthijsse2015RoleAcquired} employed an empirical STD model, STDSIM \cite{VanDerPloeg1998STDSIMMicrosimulation} to investigate HPV transmission dynamics of HPV-16 and 18. In STDSIM, each individual is assigned with both constant characteristics (e.g. sex, birth, etc.) and variable ones (e.g. heath status, number of sexual partners, etc.). The probability distributions of sex debut for men and women were defined separately, and for each individual in the simulated population, the age of first sexual contact was determined by Monte Carlo sampling. In the simulation, all events were arranged in the appropriate time order, which was called an event-driven approach. Characteristics of individuals or relationships were updated when an event happened. The occurrence of an event might generate new events, which would occur immediately or later in the simulation, or delay or cancel scheduled events. The event-driven approach in STDSIM was more efficient than a time-based approach in checking if events occurred at fixed time steps and cutting off unnecessary checks. STDSIM is the only model taking immigration and emigration into account. The demography module of the model included birth, death, immigration and emigration to represent the general population. The sexual behaviour module depicted how relationship were formed and terminated. Pairing took place on the basis of age. Individuals becoming available for a relationship and seeking for a partner were determined by probabilities, and their preferences of the relationship duration were defined by a function of age, personal interest and whether they have other sexual partners already, in which concurrent relationships were allowed. The frequency of sexual contacts within a relationship depended on the age of the man. STDSIM also modelled commercial sex, specifying the frequency of each man visiting commercial sex workers (CSW) \cite{VanDerPloeg1998STDSIMMicrosimulation, Matthijsse2015RoleAcquired}. Matthijsse et al. \cite{Matthijsse2015RoleAcquired} simulated HPV-16 and HPV-18 transmission in the network on the basis of STDSIM following the demography in the Netherlands. They 
assumed male-to-female and female-to-male transmissions sharing equal transmissibility per coital act, and the transmission parameters were calibrated to be age and type-specific. Comparing to empirical experiments on other STDs, they assumed condom uptake did not have protection against HPV, which agreed with prevention approaches mentioned in Section \ref{prevention approaches}. Explicitly, they assumed women with hysterectomy were immune to HPV infections and modelled the fraction of hysterectomy following the general age distribution. Durations of infection and acquired immunity followed a Weibull distribution that was a continuous probability distribution. They employed a grid search to choose the optimal combinations of transmission probability and distributions of acquired immunity and infection durations. As a result, duration of acquired immunity with substantial variation played a crucial part in HPV epidemiology; without it, a model could not reproduce the age distribution of HPV prevalence.

\subsection{Classification based on Prevention Strategies} \label{prevention results}

\begin{table}
\tbl{HPV Transmission Models with various Prevention Strategies. The prevention strategies could be screening, vaccination, both (hybrid) or none. Among the papers which use hybrid strategies, some hybrid strategies have the emphasis on vaccination as the main strategy and screening as back-up (vaccination-oriented hybrids), some use screening as the main strategy and vaccination as back-up (screening-oriented hybrid), and some use hybrid strategies which are `neutral' in terms of vaccination / screening orientation.}
{\begin{tabular}{m{2cm} m{3cm} m{6cm} m{4.5cm}}
        \rowcolor{qianbai}
        Prevention Strategies & Target Population & Focus & Literature\\
        \rowcolor{tubai}
        &&&Burchell et al. 2006 \cite{Burchell2006ModelingSexual}\\
        \rowcolor{tubai}
        &&&Kim et al. 2007 \cite{Kim2007MultiparameterCalibration}\\
        \rowcolor{tubai}
        &&&Bogaards et al. 2010 \cite{Bogaards2010ModelbasedEstimation}\\
        \rowcolor{tubai}
        &&&Campos et al. 2014 \cite{Campos2014UpdatedNatural}\\
        \rowcolor{tubai}
        &&&Matthijsse et al. 2015 \cite{Matthijsse2015RoleAcquired}\\
        \rowcolor{tubai}
        &&&Acedo et al. 2018 \cite{Acedo2018CalibratingLarge}\\
        \rowcolor{tubai}
        \multirow{-7}{*}{No Prevention}&&&Villanueva et al. 2019\cite{Villanueva2019CalibrationAgentbased}\\
        \rowcolor{yaluanqing}
        &&types, ages, frequencies&Campos et al. 2015 \cite{Campos2015WhenHow} \\

        \rowcolor{yaluanqing}
        &&types, following treatment&Campos et al. 2015 \cite{Campos2015ComparativeCosteffectiveness} \\
        \rowcolor{yaluanqing}
        \multirow{-3}{*}{Screening}&\multirow{-3}{*}{Females}&visits&Campos et al. 2017 \cite{Campos2017EstimatingValue} \\

        \rowcolor{xuebai}
        \cellcolor{shuangse}&& effectiveness of sex-specific immunisation & Bogaards et al. 2011 \cite{Bogaards2011SexspecificImmunization}\\
        
        \rowcolor{xuebai}
        \cellcolor{shuangse}&&vaccine simulation& Al-arydah and Smith 2011 \cite{Al-arydah2011AgestructuredModel}\\
        \rowcolor{xuebai}
        \cellcolor{shuangse}&&optimal vaccine control& Al-arydah and Malik 2017 \cite{al2017age}\\
        
        \rowcolor{xuebai}
        \cellcolor{shuangse}&&catch-up vaccination& Díez-Domingo et al. 2017 \cite{Diez-Domingo2017RandomNetwork}\\
        
        \rowcolor{xuebai}
        \cellcolor{shuangse}&\multirow{-5}{*}{Females}& cost-effectiveness of vaccinating girls and boys & Datta et al. 2019 \cite{Datta2019AssessingCosteffectiveness}\\

        \rowcolor{yingbai}
        \cellcolor{shuangse}&&vaccination coverage & Muño-Quiles et al. 2021 \cite{Munoz-Quiles2021EliminationInfections}\\ 
        
        \rowcolor{yingbai}
        \cellcolor{shuangse}&& optimal vaccine, partner age difference & Madhu and Al-arydah 2021 \cite{Madhu2021OptimalVaccine} \\
        
        \rowcolor{yingbai}
        \cellcolor{shuangse}&& impulsive vaccination & Berhe and Al-arydah 2021 \cite{Berhe2021ComputationalModeling} \\

        \rowcolor{yingbai}
        \cellcolor{shuangse}&& optimal vaccine & Al-arydah 2021 \cite{al2021two} \\
        
        \rowcolor{yingbai}
        \cellcolor{shuangse}&& effectiveness of vaccinating girls and boys & Villanueva et al. 2022 \cite{Villanueva2022MathematicalModel}\\
        
        \rowcolor{yingbai}
        \cellcolor{shuangse}\multirow{-11}{*}{Vaccination}&\multirow{-6}{*}{\makecell{Females \\and/or\\ Males}}& vaccine uptake and health education & Al-arydah 2023 \cite{Al-arydah2023MathematicalModeling} \\

        \rowcolor{danteng}
        \cellcolor{baiteng}&\cellcolor{baiteng}&& Goldhaber-Fiebert et al. 2008 \cite{Goldhaber-Fiebert2008CosteffectivenessCervical}\\ \rowcolor{danteng}\cellcolor{baiteng}&\cellcolor{baiteng}&\multirow{-2}{*}{\makecell{cost-effectiveness of screening for vaccinated\\and unvaccinated Females}} & Burger et al. 2012 \cite{Burger2012CosteffectivenessCervical}\\
        \rowcolor{tengse}
        \cellcolor{baiteng}&\cellcolor{baiteng}&& Coupé et al. 2009 \cite{Coupe2009HowScreen}\\
        \rowcolor{tengse}
        \cellcolor{baiteng}&\cellcolor{baiteng}&& Coupé et al. 2012 \cite{Coupe2012ImpactVaccine}\\
        \rowcolor{tengse}
        \cellcolor{baiteng}\multirow{-5}{*}{\makecell{\cellcolor{baiteng}Screening-\\\cellcolor{baiteng}oriented \\\cellcolor{baiteng}Hybrid}}&\cellcolor{baiteng}\multirow{-5}{*}{Females}&\multirow{-3}{*}{cost-effectiveness of screening after vaccination} & Landy et al. 2018 \cite{Landy2018WhatCervical}\\

        \rowcolor{xuebai}
        \cellcolor{yinbai}&\cellcolor{shuangse}&& Goldie et al. 2007 \cite{Goldie2007CosteffectivenessHPV}\\
        \rowcolor{xuebai}
        \cellcolor{yinbai}&\cellcolor{shuangse}&& Kim et al. 2008 \cite{Kim2008ExploringCosteffectiveness}\\
        \rowcolor{xuebai}
        \cellcolor{yinbai}&\cellcolor{shuangse}&& Olsen and Jepsen 2010 \cite{Olsen2010HumanPapillomavirus}\\
        \rowcolor{xuebai}
        \cellcolor{yinbai}&\cellcolor{shuangse}&\multirow{-4}{*}{cost-effectiveness of vaccination} & Ekwunife and Lhachimi 2017 \cite{Ekwunife2017CosteffectivenessHuman}\\
        \rowcolor{yingbai}
        \cellcolor{yinbai}&\cellcolor{shuangse} & & Bogaards et al. 2011 \cite{Bogaards2011LongtermImpact}\\
        \rowcolor{yingbai}
        \cellcolor{yinbai}&\cellcolor{shuangse} & \multirow{-2}{*}{effectiveness of vaccination} & Van de Velde et al. 2012 \cite{VanDeVelde2012PopulationlevelImpact}\\   \rowcolor{yuebai}
        \cellcolor{yinbai}&\cellcolor{shuangse}\multirow{-7}{*}{Girls} & vaccine allocation by sexual risk & Malagón et al. 2013 \cite{Malagon2013ImpactDifferential}\\
        
        \rowcolor{tubai}
        \cellcolor{yinbai}&\cellcolor{xiekeqing} & & Kim and Goldie 2008 \cite{Kim2008HealthEconomic} \\
        \rowcolor{tubai}
        \cellcolor{yinbai}&\cellcolor{xiekeqing} & \multirow{-2}{*}{cost-effectiveness of vaccination} & Coupé et al. 2009\cite{Coupe2009HPV1618}\\
        \rowcolor{su}
        \cellcolor{yinbai}&\cellcolor{xiekeqing} \multirow{-3}{*}{Females} & effectiveness of vaccination & Van de Velde et al. 2007 \cite{VanDeVelde2007ModelingHuman}\\ 
        
        \rowcolor{bailv}
        \cellcolor{yinbai}&\cellcolor{liuse} & & Kim and Goldie 2009 \cite{Kim2009CostEffectiveness}\\
        \rowcolor{bailv}
        \cellcolor{yinbai}&\cellcolor{liuse} & \multirow{-2}{*}{incremental value of adding boys } & Brisson et al. 2011 \cite{Brisson2011IncrementalImpact}\\
        \rowcolor{xiachong}
        \cellcolor{yinbai}&\cellcolor{liuse} & effectiveness of vaccination & Van de Velde et al. 2010 \cite{VanDeVelde2010UnderstandingDifferences}\\
        \rowcolor{mengcong}
        \cellcolor{yinbai}&\cellcolor{liuse} & & Laprise et al. 2014 \cite{Laprise2014ComparingCosteffectiveness}\\
        \rowcolor{mengcong}
        \cellcolor{yinbai}\multirow{-15}{*}{\makecell{\cellcolor{yinbai}Vaccination-\\\cellcolor{yinbai}oriented\\\cellcolor{yinbai}Hybrid}}&\cellcolor{liuse}\multirow{-5}{*}{Females and Males}& \multirow{-2}{*}{routine of vaccination} & Drolet et al. 2017 \cite{Drolet2017ImpactHuman}\\
        
        \rowcolor{xiangya}
        \cellcolor{gao}&\cellcolor{gao}&implementation & Goldhaber-Fiebert 2007 et al. \cite{Goldhaber-Fiebert2007ModelingHuman}\\
        \rowcolor{yudu}
        \cellcolor{gao}&\cellcolor{gao}&&Diaz et al. 2008 \cite{Diaz2008HealthEconomic} \\
        \rowcolor{yudu}
        \cellcolor{gao}&\cellcolor{gao}&\multirow{-2}{*}{cost-effectiveness}& Campos et al. 2012 \cite{Campos2012HealthEconomic}\\
        \rowcolor{baifen}
        \cellcolor{gao}\multirow{-4}{*}{Neutral Hybrid}&\cellcolor{gao}\multirow{-4}{*}{Females}&effectiveness & Jarynowski and Serafimovic 2014 \cite{Jarynowski2014StudyingPossible} \\

\end{tabular}}
\label{prevention table}
\end{table}

Literature with prevention strategies are reviewed in this section. Models without prevention strategies implemented or their primary models have been reviewed in the previous section; the details are listed in Table \ref{prevention table}. To better investigate intervention implementation, it is essential to ascertain what kind of prevention strategy is implemented in the model. We considered three cases, as mentioned before:
\begin{itemize}
    \item Screening: screening is the only implemented prevention strategy
        \item Vaccination: vaccination is the only implemented prevention strategy 
    \item Hybrid: both screening and vaccination are implemented
    \end{itemize}

\subsubsection{Screening}
WHO introduced the global strategy to accelerate elimination cervical cancer in 2020, in which 70\% of eligible women should receive HPV screening at least twice in their lifetimes \cite{web2021guideline}. The conventional screening approach has been the cytology test, introduced in the 1940s, which has reduced the mortality from cervical cancer five-fold over the past 50 years; however, due to limited resources, cytology has not been promoted in low- and middle- income countries (LMICs) \cite{Lowy2008HumanPapillomavirus, web2021guideline}. Newer technologies have provided new options for screening. VIA allows examining by naked eyes or magnifying by colposcope or camera \cite{web2021guideline}. Since cervical cancer and all precursor lesions are attributed by high-risk HPV infection, NAAT can identify possible precursors on molecular level and prevent unnecessary treatments of plausible lesions, and it is more reproducible and can be applied under all settings \cite{Lowy2008HumanPapillomavirus, web2021guideline}. In empirical experiments, NAAT detected the same amount of cases as cytology but earlier \cite{Lowy2008HumanPapillomavirus}. As a result, WHO recommends that women in general should receive HPV DNA tests every 5-10 years starting at aged 30 years; for immunoscompromised women, HPV DNA tests should be performed every 3-5 years starting at aged 25 years \cite{WorldHealthOrganization2022HumanPapillomavirus}.

In the literature, there are three publications \cite{Campos2015WhenHow, Campos2015ComparativeCosteffectiveness, Campos2017EstimatingValue} focusing on screening in LMICs, and all employed the same HPV transmission model \cite{Kim2007MultiparameterCalibration}. In the three publications \cite{Campos2015WhenHow, Campos2015ComparativeCosteffectiveness, Campos2017EstimatingValue}, types, frequencies, ages, visits and following treatments of screening were compared to evaluate different screening strategies; to analyse screening's efficiency in reducing cervical cancer risks, immediate treatment were implemented after positive screening results came out, which are shown in Table \ref{prevention table}.

Campos et al. \cite{Campos2015WhenHow} compared three types of screening tests: 1) the \textit{care}HPV (provider-collected [cervical] and self-collected [vaginal sampling]); 2) VIA; and 3) conventional cytology. Here, the \textit{care}HPV is an economically sustainable HPV DNA test introduced in places where resources are limited \cite{Cremer2016IntroducingHighRisk}. The model \cite{Campos2015WhenHow} assumed, women receiving screening during the first visit and returning for the results during the second visit, for \textit{care}HPV; eligible women with positive screening results would mainly receive cryosurgery on the same day. In the VIA group, most eligible women who were screen-positive would receive immediate cryosurgery, but a part of them would either come back for a subsequent visit or drop out; those who were not eligible would be referred to a secondary facility for further diagnosis and treatment. Cytology required multiple visits. Women needed to be screened for the first visit and come back for results during the second one. Screen-positive women would be performed diagnostic colposcopy and biopsy during the third visit and visit a fourth time for treatment if necessary. Each screening test was performed
between the ages of 25 and 50 years, at frequencies of once to three times in a lifetime, at 5 or 10 year intervals. 70\% of women were randomly selected at each target age; women who missed one screening would not be selected next time. The results suggested that \textit{care}HPV both cervical sampling and vaginal sampling were more effective and cost-effective in all settings comparing to the other two approaches. Within the age range from 30 to 45 years, screening three times could have the optimal and cost-effective performance, reducing cancer risk by 50\%.

In the same year, Campos et al. \cite{Campos2015ComparativeCosteffectiveness} investigated the health and economic value of HPV screening algorithms in El Salvador, focusing on both screening and treatment. The simulated population in the model was women aged 30-65 years. First, for HPV-based screening, \textit{care}HPV plan, women were supposed to receive screening every five years. The researchers compared two treatment plans following the positive screening results: 1) referral to colposcopy (Colposcopy Management [CM]) for women with CIN1 or higher stages at the designated hospital; 2) VIA triage followed by cryotherapy for eligible women in the clinic or colposcopy for ineligible women at the hospital (Screen and Treat [ST]). For Pap screening, women are supposed to uptake screening every two years followed by a referral to colposcopy for positive results (Pap). Here, they assumed 65\% of the simulated women receiving administered screening and the left 35\% never receiving screening. HPV-based screening demonstrated more efficiency in reducing cervical cancer risk by ~60\%, and ST was recognised as the most cost-effective strategy.

In 2017, Campos et al. \cite{Campos2017EstimatingValue} compared the 2- and 1- visit schedules of HPV DNA test to estimate the health and economic impact of reducing screening visits in LMICs. The model was calibrated to clinic data from India, Nicaragua and Uganda. In the simulation,each women were administrated to be screened three times in their lifetime starting at age 30 with five years interval, which reflected the results from their empirical study \cite{Campos2015WhenHow}. For 2-visit schedule, women would receive screening at the first visit and return for the results. the rate of loss to follow-up (LTFU) was varied from 10\% to 70\% with 10\% increment, and each level was compared to 1-visit schedule. For 1-visit schedule, women would receive results on the same day of screening. Both strategies provided same-day cryotherapy at the result visit, and the performance of the two strategies were assumed comparable. The simulation results suggested the potential of developing 1-visit screening strategy, which was also called screen-and-treat.

\subsubsection{Vaccination}
HPV vaccination can induce avid antibody or polyclonal antibody response against designated HPV genotypes, which human bodies cannot develop from infection. The current prophylactic vaccines are indicated to be administrated before individuals are exposed to HPV, as vaccination can reach the optimal efficacy before the onset of sexual activity \cite{WorldHealthOrganization2022HumanPapillomavirus}. 

In the literature, the following publications utilised computational models to evaluate different HPV vaccination strategies. According to the position paper published by WHO \cite{WorldHealthOrganization2022HumanPapillomavirus}, the primary target population of receiving HPV vaccination should be girls aged 9-14 years before their sexual debut, as prevention of cervical cancer is best achieved by immunising adolescent girls. When the vaccination coverage is high enough to achieve herd immunity, which is the threshold representing the proportion of the population that need to be immunised to the infection to sufficiently contain the further transmission \cite{Piraveenan2021OptimalGovernance}. Once the coverage of HPV vaccination in girls reaches 80\%, the herd immunity can also benefit boys. Females over 15 years old, boys, men or MSM are listed as the secondary target populations \cite{Lowy2008HumanPapillomavirus, WorldHealthOrganization2022HumanPapillomavirus}. To deliver a more comprehensive investigation on HPV vaccination in computational models, we categorised the literature according to target vaccine uptake populations, as shown in Table \ref{prevention table}.

Al-arydah and Smith \cite{Al-arydah2011AgestructuredModel}, and  Al-arydah and Malik \cite{al2017age} modelled vaccinating females only. They considered vaccination of teenage girls,  as suggested by \cite{WorldHealthOrganization2022HumanPapillomavirus}, but also modelled catch-up vaccination of females up to the age 26. In particular, Al-arydah and Smith \cite{Al-arydah2011AgestructuredModel} developed an aged-dependent two-sex deterministic SIR model to study the effect of the bi-valent HPV vaccination program targeting HPV-16 and 18. Their  model employed a time and age dependent system of partial differential equations (PDEs) for simulating HPV transmission. The vaccine in their model is given to females in two stages: the childhood vaccine for ages under 13 and the adult vaccine for ages 13-26.  It is assumed that the possibility of receiving the childhood vaccine does not depend on age as it is free of charge, while the chance of paying for the adult vaccine increases with age since adult women will have higher income and more awareness of the need for the vaccine as they grow older and gain more sexual experience. Their results suggested that  the effects of age dependency in vaccine uptake are complex, but that vaccinating a single age cohort in one gender can result in eventual control of the disease across all age groups. Meanwhile Al-arydah and Malik \cite{al2017age} introduced an age-structured SIS model based on a nonlinear partial differential equation system to study the effect of a bi-valent HPV vaccination program targeting HPV-6 and -11. HPV-6 and -11 are common low-risk HPV genotypes and can cause genital warts, and recovery cannot provide permanent immunity. Their results suggest that early and catch-up female vaccine program eliminates the infection in both male and female populations over 30 years. They introduced the optimal control to an age-dependent model based on ordinary differential equations and solved it numerically to obtain the most cost-effective method for introducing the catch-up vaccine into the population. Though the framework is essentially mathematical, the numerical solutions used computational methods.  Their results showed that, to ensure the eradication of HPV types 6 and 11, not only vaccines with high efficacy are needed, but it is also needful to  increase the proportion of early females vaccinated and use optimal catch-up vaccines. The results imply that the catch-up vaccine can be administered optimally by using the maximum possible rate at the earliest possible age right away after females become sexually active.

Díez-Domingo et al. \cite{Diez-Domingo2017RandomNetwork} devised a heterosexual contact network to simulate the HPV vaccination campaign against genital warts in Australia in 2007. In the project, two scenarios were simulated:
\begin{itemize}
    \item Scenario 1: 83\% routine vaccination coverage of girls under 14 years old plus 73\% catch-up vaccination coverage of women aged 14-26 years;
    \item Scenario 2: 73\% routine vaccination coverage of girls under 14 years old plus 52\% catch-up vaccination coverage of women aged 14-26 years.
\end{itemize}
The vaccination efficacy was assumed 96.5\%.
In the simulation results, prevalence of genital warts in both women and men dropped after the introduction of vaccination, which revealed that non-vaccinated men could also benefit from the herd immunity by vaccinating females only.

The following three publications \cite{Bogaards2011SexspecificImmunization, Datta2019AssessingCosteffectiveness, Villanueva2022MathematicalModel} considered adding boys to the immunisation program. Datta et al. \cite{Datta2019AssessingCosteffectiveness} constructed an individual-based stochastic model simulating nine types of HPV transmission in the UK. They tested three current HPV vaccines, the bivalent, quadrivalent and nonavalent vaccines, under the ``historical vaccination'' background that only girls were vaccinated during 2008-2016. In the results, vaccinating girls only was the most cost-effective strategy; adding boys to an already successful girls-only immunisation plan was not cost-effective since men could be protected by the herd immunity as well. Villanueva et al. \cite{Villanueva2022MathematicalModel} adjusted an existing heterosexual network \cite{Diez-Domingo2017RandomNetwork} with MSM implemented and simulated three vaccination scenarios:
\begin{itemize}
    \item 1. 75\% of 12- to 13-year-old girls were vaccinated;
    \item 2. 75\% of 12- to 13-year-old boys were vaccinated;
    \item 3. 75\% of 12- to 13-year-old both girls and boys (gender neutral) were vaccinated;
\end{itemize}
Similar to the results above, vaccinating boys was not cost-effective; however, MSM population could not be protected by the herd immunity when only girls were vaccinated. Vaccinating boys only could have partial control of infections in the MSM population but would put females at substantial risk. Therefore, the most effective campaign would be vaccinating both genders. Bogaards et al. \cite{Bogaards2011SexspecificImmunization} investigated the most effective sex-specific immunisation strategy and compared scenarios of vaccinating girls only, boys only and girls and boys. The objective was to vaccinate the sex with the highest prevalence of infection in order to achieve the largest reduction in population prevalence; however, the results were counterintuitive. Vaccinating the sex with the highest prevaccine prevalence yielded different results in SIS and SIR systems; vaccinating individuals at the highest infection risk was most effective in the SIR system, while in the SIS system, individuals with longer infectious periods needed prioritising. Moreover, vaccinating the sex with the higher prevaccine prevalence could not reach the expected performance if MSM were included in the model, and to reduce most prevalence in MSM, only boys could be vaccinated. However, in the conclusion, increasing vaccination coverage among girls was acknowledged as more effective since most existing HPV vaccination programs have achieved sufficient coverage with female-only vaccination.

In the following two models \cite{ Berhe2021ComputationalModeling, Munoz-Quiles2021EliminationInfections}, both females and males were vaccinated. Berhe and Al-arydah \cite{Berhe2021ComputationalModeling} proposed a novel SIVS HPV model with impulsive vaccination at regular intervals to examine the impulsive vaccine control strategy. Fractions of HPV naive girls and boys were administrated pulse vaccination before the onset of sexual activity. As a system dynamics model, the model's global stability of disease-free periodic solution agreed with the condition based on basic reproduction number  (for not having an epidemic) $R_0 \le 1$ as  mentioned in Section \ref{modelling structure method}. The results suggested that a larger pulse vaccination rate administrated over a longer period might help in eradicating the disease. In addition, the authors state that the model was limited without implementation of acquired immunity. Muñoz-Quiles et al. \cite{Munoz-Quiles2021EliminationInfections} analysed the prevalence decline in HPV infection under scenarios of 75\% and 90\% vaccination coverage in an existing sexual contact network \cite{Diez-Domingo2017RandomNetwork}. The results suggested that once the vaccination coverage in the sexual contact network reached 70\%-75\%, improving the coverage showed limited variation in prevalence reduction. Meanwhile, the results again confirmed the importance of gender-neutral vaccinating program in providing health equity for MSM.

Three further works which model HPV vaccination for both males and females are  Madhu and Al-arydah \cite{Madhu2021OptimalVaccine}, Al-arydah (2021) \cite{al2021two}, and Al-arydah (2023) \cite{Al-arydah2023MathematicalModeling}. Madhu and Al-arydah \cite{Madhu2021OptimalVaccine} introduced a two-sex age-structured cohort deterministic SIS model to describe the dynamics of HPV disease with childhood and catch-up vaccines, and discussed the impact of the age-difference of partners on the HPV prevalence and optimal vaccines. They proved that an optimal vaccine solution exists for this scenario and it  is unique. They showed that  in this scenario, 77\% coverage of childhood vaccination is essential and controls HPV in all ages in a 20-year period. They demonstrated that childhood vaccination is more efficient than adult vaccination in controlling HPV diseases in youth and adults. Meanwhile Al-arydah (2021) \cite{al2021two}  formulated a two-gender SIS  model to search for optimal childhood and catch-up vaccination strategies against HPV-6 and -11 over a 20-year period. The populations of the two genders were modelled to grow logistically. They showed that to minimise the effective  reproduction number under resource constraints in terms of HPV vaccinations,   the priority for vaccines should be given to girls. They also determined minimum threshold levels of catch-up vaccine coverage separately for both genders, which is necessary to avoid an HPV epidemic (to keep the effective reproduction number less than unity). Similarly,   Al-arydah (2023) \cite{Al-arydah2023MathematicalModeling} formulated an SIS cohort deterministic model that incorporates population behaviour and the impact of education on HPV dynamics. The goal was, specifically,  assessing health education success in controlling HPV through increasing vaccine uptake and reducing risky behaviour in individuals. The model again assumed  logistically growing populations. Al-arydah estimated the time series solution of the model for uneducated, lowly-educated and highly-educated populations. They showed that the higher the education level, the more the vaccine uptake, the less the risky behaviour, and the shorter the time needed to control HPV. They demonstrated that late education for adults is more effective than early education for children as late education is provided in tandem with vaccination programs,  and that could increase vaccine uptake for parents and their children. Essentially they argued that increasing health education is a viable low-cost strategy to support vaccines that have low efficacy.

\subsubsection{Hybrid Prevention Strategy}

To achieve the maximum prevention of cervical cancer, HPV vaccination and screening should be practised together \cite{WorldHealthOrganization2022HumanPapillomavirus}. As the primary prevention intervention, HPV vaccination can induce persistent and avid antibody response that cannot develop from virus clearance. However, the current HPV vaccines cannot cover all oncogenic HPV genotypes and prevent some cancer-related infections, and vaccination can only provide partial protection to females. Therefore, the prevention of cervical cancer needs the secondary assistance of screening \cite{Lowy2008HumanPapillomavirus, WorldHealthOrganization2022HumanPapillomavirus}. With the diagnosis of screening, responsive treatment can be administrated to defer further progress of cervical lesions. Even if the global vaccination coverage were achievable today, the role of screening as the secondary prevention of cervical cancer cannot be replaced \cite{Lowy2008HumanPapillomavirus, web2021guideline, WorldHealthOrganization2022HumanPapillomavirus}.

Considering the large quantity of HPV transmission models with hybrid prevention strategies implemented in the literature, we further categorised the related publications according to the following criteria:
\begin{itemize}
    \item Screening oriented 
    \item Vaccination oriented 
    \item Combination of vaccination and screening
\end{itemize}
The details are listed in Table \ref{prevention table}.

Five publications \cite{Goldhaber-Fiebert2008CosteffectivenessCervical, Coupe2009HowScreen, Coupe2012ImpactVaccine, Landy2018WhatCervical, Burger2012CosteffectivenessCervical} investigated cost-effectiveness of screening with different vaccination settings in the model.

Goldhaber-Fiebert et al. \cite{Goldhaber-Fiebert2008CosteffectivenessCervical} and Burger et al. \cite{Burger2012CosteffectivenessCervical} investigated the cost-effectiveness of cervical cancer screening for both vaccinated and unvaccinated females. Goldhaber-Fiebert et al. \cite{Goldhaber-Fiebert2008CosteffectivenessCervical} used an empirical individual-based stochastic model \cite{Kim2007MultiparameterCalibration} to examine screening strategies for unvaccinated older women and girls eligible for HPV vaccination. In the implementation, prevention strategies included screening along, vaccination alone and vaccination accompanied by screening, where girls were vaccinated before age 12 years. The researchers examined cytology and HPV DNA testing and varied the combination of the initial age (18, 21, or 25 years) and interval (every 1, 2, 3, or 5 years). The testing strategies included:
\begin{itemize}
    \item Cytology with HPV test triage: HPV DNA testing after cytology for suspicious cytologic results;
    \item HPV test with cytology triage: cytology after HPV DNA testing for positive HPV DNA diagnosis;
    \item Combined HPV DNA testing and cytology.
\end{itemize}
The simulated individuals were allowed to switch testing strategies once at age 25, 30 or 35 years. The results suggested the most cost-effective screening strategy for both vaccinated and unvaccinated female is to use HPV DNA testing as triage for equivocal results in young women and primary screening approach in older women basing on age. Burger et al. \cite{Burger2012CosteffectivenessCervical} employed the same primary model \cite{Kim2007MultiparameterCalibration} to find a cost-effective screening program in Norway. They compared the current cytology program with HPV DNA testing in for women who were (not) vaccinated before adolescence. Similar to Goldhaber-Fiebert et al.'s conclusion, Burger et al. \cite{Burger2012CosteffectivenessCervical} suggested to switch primary screening from cytology to HPV DNA testing in older women for cost-effectiveness.

The following three papers \cite{Coupe2009HowScreen, Landy2018WhatCervical} investigated cost-effectiveness of screening after vaccination. Coupé et al. \cite{Coupe2009HowScreen} applied a cohort-based stochastic model \cite{Coupe2009HPV1618} to evaluate the optimal approach to screen women after vaccination. They compared cytology and HPV DNA test as primary screening instrument. As results, screening five times with HPV DNA test or seven times with cytology between 30 and 60 years were comparable in cost-effectiveness. Later, Coupé et al. \cite{Coupe2012ImpactVaccine} compared screening scenarios after different types of vaccination. The first one is a bivalent HPV vaccine with partial cross-protection against HPV-31, 33, 45 and 58; the second one is a broad spectrum vaccine against HPV-5 to 13. In the simulation, screening once per lifetime was more cost-effective than receiving a broad spectrum vaccination. For females receiving the bivalent vaccine, they were suggested to take screening four times per lifetime, while for the ones taking broad spectrum vaccination, one screen was more cost-effective. Landy et al. \cite{Landy2018WhatCervical} proposed a microsimulation model to determine the appropriate screening routine for females after vaccination. The simulation results suggested that females vaccinated against HPV-16 and 18 needed three screening tests during their lifetime, females vaccinated against HPV-16, 18, 31, 33, 45, 52 and 58 required twice, and the unvaccinated females should receive seven lifetime screens.

The majority of papers with hybrid prevention strategy implemented are vaccination-oriented, as shown in Table  \ref{prevention table}. To better deliver the implementation in the models, the related publications are divided into the following subcategories based on vaccinated population:
\begin{itemize}
    \item Girls
    \item Females (girls and young women)
    \item Females and males
\end{itemize}
In the models \cite{Goldie2007CosteffectivenessHPV, Kim2008ExploringCosteffectiveness, Olsen2010HumanPapillomavirus, Ekwunife2017CosteffectivenessHuman, Bogaards2011LongtermImpact, VanDeVelde2012PopulationlevelImpact, Malagon2013ImpactDifferential}, adolescent girls were prioritised to be vaccinated as guided by WHO \cite{WorldHealthOrganization2022HumanPapillomavirus}; on the premise of ensuring the vaccination coverage in girls, young women \cite{Kim2008HealthEconomic, Coupe2009HPV1618, VanDeVelde2007ModelingHuman} and boys or males \cite{Kim2009CostEffectiveness, Brisson2011IncrementalImpact, VanDeVelde2010UnderstandingDifferences, Laprise2014ComparingCosteffectiveness, Drolet2017ImpactHuman} were included.

Goldie et al. \cite{Goldie2007CosteffectivenessHPV} and Kim et al. \cite{Kim2008ExploringCosteffectiveness} used an individual-based simulation model \cite{Kim2007MultiparameterCalibration} to explore the health and economic impact of HPV vaccination in Brazil and Vietnam. In the model \cite{Kim2007MultiparameterCalibration}, girls were vaccinated, and females received cervical cancer screening two or three time per lifetime since they were 30. HPV DNA test and cytology were compared as the primary screening approach. Both Goldie et al. \cite{Goldie2007CosteffectivenessHPV} and Kim et al. \cite{Kim2008ExploringCosteffectiveness} suggested the combination of vaccination and screening could achieve the optimal reduction in cervical cancer risk; however, for LMICs, without assistance, they might not afford the vaccines, and screening only ended up the most cost-effective approach. Olsen and Jepsen \cite{Olsen2010HumanPapillomavirus} constructed an agent-based heterosexual contact model to analyse the incremental impact of introducing HPV vaccination compared to screening alone in Denmark. In the model, the screening program covered 70\% of women aged 23-59 years in Denmark, who received screening every three years. Quadrivalent HPV vaccination was allocated to 70\% of 12-year-old girls without catch-up program or booster dose in the analysis. The results suggested that though the introduction of quadrivalent vaccination would incur extra cost, the risk of cervical cancer and cost of treatment would be reduced. Ekwunife and Lhachimi \cite{Ekwunife2017CosteffectivenessHuman} remodelled a microsimulation framework to estimate the cost-effectiveness of introducing HPV vaccination to Nigeria. 
They conducted scenarios in the simulation including:
\begin{itemize}
    \item Current opportunistic cervical cancer screening (CS): VIA once per lifetime with 8.7\% coverage;
    \item National cervical cancer screening (NS): VIA once per lifetime at age 30 for each woman;
    \item National HPV vaccination with opportunistic screening (CS + NV): two doses of bivalent HPV vaccines for preadolescent girls at age 12 in addition to opportunistic screening;
    \item National screening and national HPV vaccination (NS + NV): two doses of bivalent HPV vaccines for preadolescent girls at age 12 and VIA for women at age 30 once per lifetime
\end{itemize}
The CS + NV strategy was the only cost-effective and robust option when the cost of vaccines were relatively low enough, which was a problem for LMICs.

Bogaards et al. \cite{Bogaards2011LongtermImpact} studied the heterosexual transmission of HPV infection \cite{Bogaards2010ModelbasedEstimation} to disentangle the direct and indirect effects of HPV transmission. They implemented the current cytologic screening program in the Netherlands in the model and simulated vaccination uptake among girls. The results demonstrated that the reduction in indirectly averted cancer cases peaked at 50-70 \% vaccination coverage, which means vaccinating preadolescent girls could reduce transmission of vaccine-preventable HPV infection and lower infection rates in unvaccinated women. Van de Velde et al. \cite{VanDeVelde2012PopulationlevelImpact} employed the HPV-ADVISE model \cite{VanDeVelde2010UnderstandingDifferences} to compare population-level impact of the bivalent, quadrivalent and nonavalent HPV vaccines. In the model \cite{VanDeVelde2012PopulationlevelImpact}, both vaccination and screening were implemented as potential heath-seeking behaviour. There were five levels of screening behaviours representing the average interval between two routine cytology test when there was no equivocal result, ranging from every 1.25 years (Level 0) to never (Level 4); responsive treatment approaches were implemented to corresponding test results. Female's screening rates depended on screening behaviour level, previous screening test results and age. Sensitive analysis was conducted on the three different types of vaccines. In the results, the three types of vaccines displayed different advantages. The bivalent vaccine was more effective at preventing high level lesions and cervical cancer in the long run, while the quadrivalent vaccine substantially reduced anogenital warts soon after introduction. They also suggested switching to the nonavalent vaccine could further decrease the risk of precancerous lesions and invasive cervical cancer. 

Malagón et al. \cite{Malagon2013ImpactDifferential} calibrated the HPV-ADVISE model \cite{VanDeVelde2010UnderstandingDifferences} to Canadian epidemiological data. The simulated population was stratified into four levels according to sexual activity (low = L0, high = L3), where the most sexual active groups represented a minority of the population but accounted for almost half of the prevalence. All girls were vaccinated at age 12 years with 100\% efficacy against HPV-16 and 18, and the protection will last 20 years on average in the base case scenario. In the simulation, one-way sensitivity analysis was performed on vaccine uptake by sexual activity level. The results implied that the impact of uniform vaccination coverage on population effectiveness might be limited since the coverage was not high enough to produce herd immunity in high risk group; HPV vaccine uptake should be distributed heterogeneously according to risk of HPV-16 and 18 infection.

The following three models \cite{Kim2008HealthEconomic, Coupe2009HPV1618, VanDeVelde2007ModelingHuman} considered the inclusion of young women in the immunisation program. 

Kim and Goldie \cite{Kim2008HealthEconomic} and Coupé et al. \cite{Coupe2009HPV1618} investigated the cost-effectiveness of HPV vaccination in females. Kim and Goldie \cite{Kim2008HealthEconomic} adopted a dynamic heterosexual model \cite{Kim2007MultiparameterCalibration} to study the health and economic implications of HPV vaccination in the U.S. In the model, all prevention strategies contained screening, which was implemented following the local reported pattern. The researcher allocated catch-up vaccination from age 13 to 18, 21, or 26 in addition to routine vaccination at age 12 for girls and young women. To be noted, in the model, they assumed that unvaccinated females were also less likely to receive screening. The results suggested that the cost-effectiveness could be optimised by improving vaccination coverage in preadolescent girls and age threshold of catch-up vaccination to 18 or 21 years of age. The health impact of HPV vaccination also depended on the immunity duration, and screening policies needed revising as the introduction of vaccination. Coupé et al. \cite{Coupe2009HPV1618} developed a simulation model accommodating both screening and vaccination. In the model, females aged 30-60 were invited to cytology every five years, and 85\% 12-year-old girls were assumed to receive three doses of HPV vaccines. In addition, they separately investigated the impact of administrating catch-up vaccination at 30 years of age. The following scenarios were conducted:
\begin{itemize}
    \item No prevention
    \item Screening only
    \item Vaccination only
    \item Combination of screening and vaccination
\end{itemize}
The simulation results confirmed the health impact of vaccination in reducing risk of cervical cancer and supported that vaccination should be introduced along with screening to achieve better effectiveness. 

Van de Velde et al. \cite{VanDeVelde2007ModelingHuman} developed a stochastic model of natural history of HPV infection to predict the impact of quadrivalent HPV vaccination and measure parameter uncertainty. The probabilities of being screened was set as age-specific, and the sensitivity of test was lesion-specific. Sensitive analysis was performed on the quadrivalent HPV vaccination regarding to factors as follows:
\begin{itemize}
    \item Vaccine efficacy: 85\%, 90\%, 95\%, 100\%
    \item Vaccine duration: life, 30 years, 30 years plus booster at age 30 years, 30 years plus booster every 10 years
    \item Vaccinated age: 12, 15, 20, 30 years
\end{itemize}
As results, the effectiveness of vaccination might be influenced by vaccinated age, vaccine efficacy and immunity duration; vaccinating girls at 12 years of age with high efficacy and lifetime protection could reduce considerate risk of HPV infection.

Five other publications \cite{Kim2009CostEffectiveness, Brisson2011IncrementalImpact, VanDeVelde2010UnderstandingDifferences, Laprise2014ComparingCosteffectiveness, Drolet2017ImpactHuman} accounted males in the target vaccinated population.

First, Kim and Goldie \cite{Kim2009CostEffectiveness} and Brisson et al. \cite{Brisson2011IncrementalImpact} investigated the incremental value of including boys in the immunisation program. Kim and Goldie utilised the empirical model \cite{Kim2008HealthEconomic} introduced above, in which both girls and boys aged 12 years were vaccinated as inputs in the context of cervical cancer screening. Vaccinating both girls and boys were compared to vaccinating girls only, and the cost of including boys outweighed the benefits it would bring. As a result, the inclusion of boys needed iteratively revising. Brisson et al. \cite{Brisson2011IncrementalImpact} simulated vaccinating 12-year-old boys besides girls in the HPV-ADVISE model \cite{VanDeVelde2010UnderstandingDifferences}, where the vaccination could provide up to 20-year protection with 99\% efficacy and 70\% coverage. There was an incremental reduction in HPV incidence over 70 years; however, as the vaccination coverage increased in girls, the incremental benefit of vaccinating boys became limited.

Van de Velde et al. \cite{VanDeVelde2010UnderstandingDifferences} proposed the HPV-ADVISE model to explore factors influencing predicted effectiveness of HPV vaccination, where screening was implemented as potential behaviour in the biological process. Three different vaccination strategies were compared: 1) girls only, 2) girls and boys, and 3) girls with a 1-year catch-up program up to age 18 year, in which characteristics of vaccination were varied, including efficacy, duration, coverage and uptake age. The results implied that the prediction of HPV vaccine effectiveness was sensitive to vaccine efficacy, duration and coverage. Both adding boys and catch-up vaccination could help with HPV infection elimination.

Laprise et al. \cite{Laprise2014ComparingCosteffectiveness} and Drolet et al. \cite{Drolet2017ImpactHuman} both employed the HPV-ADVISE model \cite{VanDeVelde2010UnderstandingDifferences}. Laprise et al. \cite{Laprise2014ComparingCosteffectiveness} performed non-inferiority trial on multiple vaccination strategies:
\begin{itemize}
    \item No vaccination vs. two-dose girls-only;
    \item Two-dose girls-only vs. three-dose girls-only;
    \item Two-dose girls-only vs. two-dose girls and boys;
    \item Three-dose girls and boys vs. two-dose girls and boys or three-dose girls-only.
\end{itemize}
In the simulation, the efficacy duration of two doses was varied between 10 years and lifelong, and that of three doses was varied between 20 years and lifelong. Under most scenarios, vaccinating boys was not cost-effective unless the cost for boys was substantially lower than that for girl; if the protection duration of two doses reached 10 years, two-dose routing was likely to be cost-effective. Drolet et al. \cite{Drolet2017ImpactHuman} estimated incremental effectiveness of administrating catch-up quadrivalent vaccine to females and males in Australia. Simulations were performed from no vaccination to routine vaccination, routine plus catch-up 18 years and routine plus 26 years with vaccinating girls only or girls and boys. The simulation results exemplified that catch-up vaccination produced rapid declines in prevalence and strong herd immunity effects along with high coverage of routine vaccination. The added benefits brought by catch-up vaccination could last for 20-70 years.

The following publications \cite{Goldhaber-Fiebert2007ModelingHuman, Diaz2008HealthEconomic, Campos2012HealthEconomic, Coupe2012ImpactVaccine, Jarynowski2014StudyingPossible} explored the combination of vaccination and screening as prevention interventions in HPV transmission models. 

Goldhaber-Fiebert et al. \cite{Goldhaber-Fiebert2007ModelingHuman}, Diaz et al. \cite{Diaz2008HealthEconomic} and Campos et al. \cite{Campos2012HealthEconomic} used the same individual-based stochastic model developed by Kim et al. \cite{Kim2007MultiparameterCalibration}. Goldhaber-Fiebert et al. \cite{Goldhaber-Fiebert2007ModelingHuman} included screening and vaccination implementation in the primary model \cite{Kim2007MultiparameterCalibration}. Simulated prevention interventions contained vaccination only, screening only annually or biennially, and combination of screening and vaccination. Practising screening or vaccination individually could reduce lifetime risk of cancer; vaccination combined with screening every five years achieved the highest reduction in the results. 

Diaz et al. \cite{Diaz2008HealthEconomic} and Campos et al. \cite{Campos2012HealthEconomic} examined health and economic impact of cervical cancer screening and bivalent HPV vaccination in India and Eastern Africa. Preadolescent girls under age 12 were vaccinated, women over age 20 got screened, and the combination of vaccination and screening were simulated in the model. The screening types included cytology, VIA and HPV DNA test, containing 1-3 visits for 1-3 times per lifetime at age range 35-45. Both publications \cite{Diaz2008HealthEconomic, Campos2012HealthEconomic} agreed that vaccination combined with HPV DNA testing three times during lifetime was the most cost-effective prevention strategy; however, as LMICs, the coverage of vaccination being high enough to provide long-term population-level protection was only achievable if the cost of HPV vaccine was economic enough.

Jarynowski and Serafimovic \cite{Jarynowski2014StudyingPossible} constructed an epidemic model and analysed impact of potential epidemiological control strategies in Poland in the future 25 years. They mimicked the actual screening situations in Poland as follows:
\begin{itemize}
    \item Until 2005: opportunistic screening without regular routine;
    \item 2006-2015: regular screening was introduced;
    \item 2016-2039: every 10 years, 1/3 women would do screening every three years.
\end{itemize}
As for vaccination, girls who were 14 years old were administered mandatory vaccines, and among girls aged 20 to 24, a subpopulation of 5\%, opted to have catch-up vaccination. The results indicated that vaccination along with screening would be effective in controlling cervical cancer risk in Poland.

\section{Discussion} \label{discussion}
A number of empirical reviews have previously examined mathematical HPV models and evaluated related intervention methods \cite{Cuzick2000SystematicReview, Dasbach2006MathematicalModels, Kim2008ModelingCervical, Goldie2008MathematicalModels, Jit2011HumanPapillomavirus, Canfell2012ModelingPreventative, Seto2012CostEffectiveness, Iskandar2022MathematicalModels}. Most of these reviews however, have addressed mathematical models \cite{Dasbach2006MathematicalModels, Kim2008ModelingCervical, Goldie2008MathematicalModels, Iskandar2022MathematicalModels}, discussed the economic and financial considerations of intervention methods \cite{Seto2012CostEffectiveness}, or generally reviewed the existing intervention strategies from a medical and psychosocial perspective \cite{Cuzick2000SystematicReview, Jit2011HumanPapillomavirus, Canfell2012ModelingPreventative}. A systematic review focusing on the computational models that can be used for HPV spread and prevention modelling has been lacking. In contrast, this paper concentrates on state-of-the-art computational HPV transmission models, aiming to deliver a comprehensive review on the related literature, investigate current research trends, and identify existing computational methods which could be useful but hitherto have not been used in modelling HPV infection dynamics and prevention strategies. The review covers a range of computational models, including system dynamics models, network growth-based models and agent-based models.

Computational HPV models serve as practical and sophisticated tools for predicting the final epidemic magnitude and evaluating optimal intervention strategies in the context of HPV infection. Computational models are able to capture a large range of attributes belonging to each individual, and thus can contain higher fidelity  and heterogeneity compared to mathematical models. The advantages and disadvantages of the three types of computational HPV models, namely system dynamics models, network growth-based models and agent-based models, could be summarised as follows.

The system dynamics models can capture the high level of heterogeneity of the simulated population. The ability to depict social heterogeneity and stochastic interactions, and ability to simulate detailed transmission models allow system dynamics models to capture the contagion dissemination process with a high level of fealty and accuracy. Nevertheless, system dynamics models are computationally inexpensive compared to network growth-based and agent-based models, making them a preferred choice during real-world epidemic events to make quick predictions. The main weakness of system dynamics models is the lack of explicitly modelled contact networks, which make it challenging to keep track of HPV transmission routes. 

By contrast, agent-based models are able to thoroughly mimic HPV transmission because individuals are modelled as `agents' and their interactions are thus explicitly modelled. This allows for a high level of fidelity and allows agent-based models to capture individual attributes at micro level and simulate infection transmission spatially and causally. With the assistance of census data and GIS, large-scale agent-based models can be employed to study spatio-temporal percolation of HPV infection. An agent-based modelling approach also gives higher levels of reusability, modifiability, flexibility and granularity. However, the approach is computationally intensive, requiring high performance computing machinery, large data storage capacity, and typically presupposes the availability of a large-amount of high fidelity data such as census data.

As the third approach, the network growth-based modelling is an approach which allows for tracking infection transmission routes, while not requiring as much computing power as an agent-based approach. In this approach, the accurate modelling of individual attributes of people is compromised to achieve simpler yet accurate contact modelling. The transmission dynamics is typically captured temporally and topologically, but not spatially. An emphasis of the underlying network topology is a feature of this modelling approach, which often sufficiently captures interaction patterns, but not necessarily the individual attributes of people which drive these patterns. In such models, the diversity of interactions is captured by the heterogeneity of the topology of the network, often represented as the degree distribution and other topological metrics of the network. Infection dynamics, including the probability of getting infected, then depend on such topological attributes as node degree. Typically the network is constructed to include all sexual contacts over a period of time, and therefore static, but the construction of longitudinal networks is possible \cite{Carvalho2012EpidemicsScenarios, Diez-Domingo2017RandomNetwork}. Simulation of infection transmission is usually then conducted on such a static network, and time-dependent partnership dissolution is typically implemented: once a relationship is formed, sexual contact frequency can change as age increases but the relationship is theoretically considered permanent, even if contact frequency approaches zero. It is however possible to develop and use network growth models which capture the temporal dynamics of link formation and deletion, representing relationship formation and discontinuation among people \cite{Wang2024SeCoNetHeterosexual}.

It should be noted that, as the primary protection against HPV infection, vaccination is the key to achieve herd immunity in the context of HPV spread. Females typically bear higher risk of acquiring HPV and developing HPV-related complications, especially cervical cancer. Therefore, the majority of vaccination uptake models that have been studied here concentrate on vaccinating females, especially adolescent girls who are yet to make their sexual debut and can get full protection from the vaccination \cite{WorldHealthOrganization2022HumanPapillomavirus}.  In addition, it has been observed that increasing the vaccination coverage in subpopulations of girls who are potentially  more sexually active may substantially improve herd immunity \cite{Malagon2013ImpactDifferential}, which is a useful insight to vaccination decision makers. Meanwhile, it can be observed that a number of studies \cite{VanDeVelde2007ModelingHuman, Kim2008HealthEconomic, Coupe2009HPV1618, Diez-Domingo2017RandomNetwork, Drolet2017ImpactHuman} have focused on catch-up vaccination, which expand the target population (in terms of age, for example) to improve vaccination coverage; however, the implemented age thresholds are outdated in many instances, as many countries have in recent years further expanded age limits, allowing vaccination for people up to 45 years old in some cases \cite{WorldHealthOrganization2022HumanPapillomavirus}. 

Nevertheless, there has also been considerable interest in vaccinating males. The research focusing on modelling the vaccination of males against HPV has switched, in recent years, from modelling cost-effectiveness \cite{Kim2009CostEffectiveness, Datta2019AssessingCosteffectiveness} in vaccinating men who have heterosexual orientation, to modelling the vaccination of MSM populations \cite{Brisson2011IncrementalImpact, Bogaards2011SexspecificImmunization, Munoz-Quiles2021EliminationInfections, Villanueva2022MathematicalModel}. The rationale for this seems to be that,  with high coverage of vaccination in girls, both unvaccinated females and heterosexual males can benefit from the herd immunity; however, there is no such indirect protection for MSM population \cite{Munoz-Quiles2021EliminationInfections, Villanueva2022MathematicalModel}. It has been observed in literature that MSM population exists as a reservoir of HPV infections \cite{Villanueva2022MathematicalModel}, and receptive anal intercourse is a crucial transmission route for HPV infection in homosexual and bisexual men \cite{Burchell2006EpidemiologyTransmission, Villanueva2022MathematicalModel}. There is limited research on studying male-to-male HPV transmission \cite{Munoz-Quiles2021EliminationInfections, Villanueva2022MathematicalModel}, and the physiology of transmission is also different from that in heterosexual contacts \cite{Burchell2006EpidemiologyTransmission}. Because cervical cancer and related precancerous lesions are the most serious HPV-related complications, MSM populations and their unique transmission mechanism have perhaps not received sufficient attention in empirical studies. However, when HPV is eliminated in heterosexual populations, it could potentially remain in MSM populations and later spread back to heterosexual populations, and modelling vaccination efforts in MSM populations is  therefore crucial for understanding HPV dynamics in the society as a whole. Indeed, (teenage) boys and MSM populations have been included in some countries for HPV vaccination programs for this very reason \cite{WorldHealthOrganization2022HumanPapillomavirus, AustralianGovernmentDepartmentOfHealthAndAgedCare2023HPVHuman}. We observed, as mentioned above, that an increasing number of studies are therefore beginning to focus on HPV vaccination programs among males.

In addition, empirical studies \cite{Laprise2014ComparingCosteffectiveness, Berhe2021ComputationalModeling} have considered the number of doses needed. In many countries, according to the licenses of current prophylactic vaccines as listed in Table \ref{vaccines table}, girls and boys aged 9-14 years are supposed to receive two doses, while the ones above 14 years of age have to receive 3 doses \cite{WorldHealthOrganization2022HumanPapillomavirus}.  However, as vaccine efficacy increases, some countries such as Australia, have stipulated that people between 12-26 years of age only need to receive one dose of nonavalent vaccine \cite{AustralianGovernmentDepartmentOfHealthAndAgedCare2023HPVHuman}. We have reviewed literature which discusses this aspect.

A number of studies that we reviewed also looked at screening efforts. To achieve optimal reduction in the risk of cervical cancer, the empirical studies suggested that vaccination should be conducted along with screening \cite{Goldhaber-Fiebert2007ModelingHuman, Coupe2009HPV1618, Jarynowski2014StudyingPossible, Ekwunife2017CosteffectivenessHuman, Olsen2010HumanPapillomavirus, Goldie2007CosteffectivenessHPV, Kim2008HealthEconomic}. However, it is noted in literature reviewed here that this is not applicable in all settings. In particular, due to limited resources, LMICs cannot simply replicate the prevention strategies in developed countries. Although the combination of vaccination and screening could achieve the maximum reduction in cervical cancer risk, the cost-effectiveness of introducing vaccination to LMICs remains concerning. Nevertheless, compared to other screening approaches, HPV DNA test is shown to be more affordable and adaptable to different settings with a relatively lower cost and high sensitivity \cite{Campos2015WhenHow, Campos2015ComparativeCosteffectiveness, web2021guideline}. Therefore, it is observed that some countries which cannot afford the cost of vaccination choose to use screening only as the most cost-effective prevention strategy   \cite{Campos2012HealthEconomic, Diaz2008HealthEconomic, Ekwunife2017CosteffectivenessHuman, Goldie2007CosteffectivenessHPV, Kim2008ExploringCosteffectiveness}. Current screening approaches mentioned in the literature include cytology, HPV DNA test and VIA , and these have been repeatedly evaluated in computational models \cite{Goldhaber-Fiebert2008CosteffectivenessCervical, Coupe2009HowScreen, Burger2012CosteffectivenessCervical, Campos2015WhenHow, Campos2015ComparativeCosteffectiveness}. It is also noteworthy that some of these models have focused on the `screen and treat' approach \cite{Campos2015WhenHow, Campos2015ComparativeCosteffectiveness}, while others went further by proposing a `screen, triage and treat' approach  \cite{Goldhaber-Fiebert2008CosteffectivenessCervical, Coupe2009HowScreen}. The simulated results of empirical studies generally agree with the current screening routine which stipulates that  females should start screening at age 30, and the frequency and following treatment should be revised according to the type of screening approach and the stage of HPV infection. 

A few other observations can be made based on the analysed literature. One is that the modelling of acquired immunity plays a vital role in simulating HPV transmission and evaluation of prevention strategies. In general, immunity for an infection is acquired through vaccination or exposure to the pathogen \cite{Piraveenan2021OptimalGovernance}. However, in the case of HPV, infection typically does not induce avid antibody response, so the acquired immunity against HPV can only come from vaccination \cite{WorldHealthOrganization2022HumanPapillomavirus}. Regardless, some compartmental models have indeed represented natural acquired immunity in `recovered' compartments \cite{Matthijsse2015RoleAcquired, Jarynowski2014StudyingPossible, Bogaards2010ModelbasedEstimation, VanDeVelde2007ModelingHuman, Datta2019AssessingCosteffectiveness}, and the validity of this is debatable. Nevertheless, the existence of even moderate levels of acquired immunity can affect the selection of target populations or individuals to administrate vaccination \cite{Bogaards2011SexspecificImmunization}. As for vaccination, vaccine-induced immunity can wane with time, but modelling this waning rate has been difficult due to a lack of clinical data, and in any case dependent on the type of vaccine used.

The modelling of intervention strategies against HPV need constant revision \cite{Carter2011HPVInfection}, as new prevention approaches such as the inclusion of boys in the immunisation program, the improvement of age threshold for catch-up vaccination, the change of vaccination routine, etc. have been introduced in recent years \cite{WorldHealthOrganization2022HumanPapillomavirus, AustralianGovernmentDepartmentOfHealthAndAgedCare2023HPVHuman}. Although the current prophylactic HPV vaccines do not cover all types, the valence has increased from two to nine \cite{WorldHealthOrganization2022HumanPapillomavirus}. Therefore, in the near future, the improvement of vaccination coverage and valence will necessitate further revision of current screening strategies. It is also possible that as vaccination saturation approaches the threshold needed for herd immunity, the importance attached to screening might drastically reduce, similar to the reverse transcriptase polymerase chain reaction (RT-PCR) tests \cite{Garcia-Finana2021RapidAntigen} for SARS-CoV-2 losing significance as COVID-19 vaccination uptake reached saturation in communities.

Finally, it is worth noting that there are gaps in reviewed literature in terms of the computational methods used. Namely, a number of computational methods used in the contexts of other diseases or epidemics have not been utilised in literature addressing HPV. Game theory \cite{Bauch2004VaccinationTheory, Piraveenan2021OptimalGovernance, Chang2020GameTheoretic} and linear programming \cite{Mbah2011ResourceAllocation, Kasaie2013SimulationOptimization, Sambaturu2020DesigningEffective, Biswas2022DesigningOptimal} are two prominent examples of this. Considering the use of game theory, it could be noted that the implementation of HPV prevention strategies typically do not consider individuals' willingness to adopt or cooperate with the intervention. During an epidemic, individuals do not decide independently whether or not to follow an intervention policy, and interdependent decision making takes place \cite{Bauch2004VaccinationTheory, Piraveenan2021OptimalGovernance, Chang2020GameTheoretic}. Game theory has been applied in modelling interventions uptake to contain the spread of diseases such as  influenza, HIV infection and COVID-19, among others. In such cases, a `player' in game theoretic parlance denotes an individual who attempts to decide whether to take a particular vaccination. Each ``player'' evaluates factors relevant to the decision including the risk of infection, severity of the disease, cost of the vaccination and the existing vaccination coverage to make the decision optimising their own profit or benefit, called ``pay-off'' or ``utility'' of a ``player''. Such a decision making process may involve self-learning (aspiration game) and social learning (imitation game), where ``players'' update their strategies relying on their knowledge and the interactions with their neighbours \cite{Piraveenan2021OptimalGovernance, Chang2020GameTheoretic}. Game theory can explicitly distinguish between  modelling compulsory vaccination and volunteer vaccination uptake. Thus, the inclusion of game theory in computational HPV transmission models will be useful in evaluating prevention options. Similarly, it could be noted that In the context of HPV transmission, the existing literature still focuses on analysing specific intervention strategies, instead of treating the vaccination program as an optimisation problem which is supposed to maximise `public good' \cite{Kim2008HealthEconomic, VanDeVelde2012PopulationlevelImpact, Diez-Domingo2017RandomNetwork, Datta2019AssessingCosteffectiveness, Villanueva2022MathematicalModel}. Linear programming is an approach which can be used to construct and analyse such an optimisation problem, especially under resource scarcity \cite{Mbah2011ResourceAllocation, Kasaie2013SimulationOptimization, Sambaturu2020DesigningEffective, Biswas2022DesigningOptimal}. Thus, linear programming can be applied to improve HPV vaccination allocation in order to accelerate HPV elimination and improve cost-effectiveness.

\section{Conclusion} \label{conclusion}

In this paper, we presented a comprehensive review of state-of-the-art computational approaches used for HPV transmission and prevention modelling. We performed a systematic search based on the PRISMA methodology to select the papers to review, eventually selecting 45 papers out of an initial pool of 10497 publications. We classified existing HPV disease dynamics models into three classes: i) system dynamics models, ii) network growth-based models, and iii) agent-based models. System dynamics models are further classified into cohort-based models and individual-based models. In terms of prevention strategies, we classified  papers as those that use i) screening, ii) vaccination and iii) hybrid prevention strategies.

Computational epidemiology provides an interdisciplinary tool for simulating HPV transmission and exploring optional public health interventions. Compared to mathematical models, computational models can capture individual and population attributes with a higher level of fidelity and granularity. The existing literature highlights that a range of computational methods are already used effectively to model HPV dynamics and prevention methods. However, we also identified gaps in the literature, compared in particular to computational methods used to capture disease dynamics of other infectious diseases, such as COVID-19, Influenza, and AIDS. Game theoretic methods and linear programming are two prominent examples of computational methods not yet used in modelling HPV dynamics and prevention, but used extensively in the context of other diseases.

We identified that in general, the current optimal strategy of eliminating HPV infection and related diseases as identified in the literature is to practice vaccination along with screening. Analysis of literature further highlights that adolescent girls should be prioritised in receiving HPV vaccination. It could be noted that the relative cost-effectiveness of various prevention strategies is context-dependent: in LMICs, practising screening-only may be more cost-effective than combining vaccination with screening due to the shortage of resources, while in developed countries, a hybrid strategy of screening and vaccination may give optimal results. Thus, the optimal prevention strategies against HPV and related diseases need constant revising based on the resources available.

It could be predicted that in the future, computational models may begin to incorporate census and GIS data to simulate and capture spatio-temporal transmission patterns of HPV. Moreover, the models may begin to take individual willingness of intervention uptake into account to evaluate the impact of individuals' decisions on the global HPV transmission and prevention. In short, while computational models are increasingly being utilised to model and analyse HPV transmission and prevention strategies, there remains a lot of further potential to apply computational epidemiology and all the tools and methods available to it in the modelling, analysis, and prevention of HPV.

\section*{Disclosure statement}

No potential conflict of interest was reported by the author(s).

\section*{Funding}

The author(s) reported there is no funding associated with the work featured in this article.

\bibliographystyle{tfq}
\bibliography{jbd}

\end{document}